\documentclass[preprint,12pt]{elsarticle}

\usepackage{amssymb}
\usepackage{amsmath}
\usepackage{graphicx}
\usepackage{ifthen}
\usepackage{epsfig}
\usepackage{array}
\usepackage{epic}
\usepackage{color}
\usepackage[table]{xcolor}
\usepackage{hyperref}
\usepackage{xspace}

\definecolor{darkgreen}{rgb}{0,0.44,0}
\definecolor{darkred}{rgb}{0.44,0,0}
\definecolor{darkblue}{rgb}{0,0,0.44}

\usepackage{arydshln}

\newcolumntype{I}{!{\vrule width 1.5pt}}

\newlength\savedwidth 

\newcommand\whline{\noalign{\global\savedwidth\arrayrulewidth 
                            \global\arrayrulewidth 1.5pt}%
           \hline 
           \noalign{\global\arrayrulewidth\savedwidth}}

\newcommand{\FlaTwoByTwo}[4]{
\left(
\begin{array}{c I c}
#1 & #2 \\ \whline 
#3 & #4 
\end{array}
\right) 
}

\newcommand{\FlaOneByTwo}[2]{
\left(
\begin{array}{c I c}
#1 & #2
\end{array}
\right)
}

\newcommand{\FlaThreeByThreeTL}[9]{
\left(
\begin{array}{c | c I c}
#1 & #2 & #3 \\ \hline
#4 & #5 & #6 \\ \whline
#7 & #8 & #9
\end{array}
\right)
}

\newcommand{\FlaThreeByThreeBR}[9]{
\left(
\begin{array}{c I c | c}
#1 & #2 & #3 \\ \whline 
#4 & #5 & #6 \\ \hline 
#7 & #8 & #9 
\end{array}
\right) 
}

\newcommand{\operation}{}

\newcommand{\routinename}{}

\newcommand{\precondition}{~}

\newcommand{\postcondition}{~}

\newcommand{\invariant}{~}

\newcommand{\guard}{~}

\newcommand{\partitionings}{~}

\newcommand{\partitionsizes}{~}

\newcommand{\blocksize}{blank}

\newcommand{\repartitionings}{~}

\newcommand{\repartitionsizes}{~}

\newcommand{\moveboundaries}{~}

\newcommand{\beforeupdate}{~}

\newcommand{\afterupdate}{~}

\newcommand{\update}{~}

\newcommand{\resetsteps}{

\renewcommand{\operation}{\phantom{[A] = op( A )}}

\renewcommand{\routinename}{\operation}

\renewcommand{\precondition}{\phantom{A = \widehat A}}

\renewcommand{\postcondition}{\phantom{A = \widehat A}}

\renewcommand{\invariant}{\phantom{ \FlaTwoByTwo{A_{TL}}{A_{TR}}{A_{BL}}{A_{BR}} =
		\FlaTwoByTwo{A_{TL}}{A_{TR}}{A_{BL}}{A_{BR}}
		\wedge
		\FlaTwoByTwo{A_{TL}}{A_{TR}}{A_{BL}}{A_{BR}} =
		\FlaTwoByTwo{A_{TL}}{A_{TR}}{A_{BL}}{A_{BR}}~~~~~~~
		}}

\renewcommand{\blocksize}{blank}

\renewcommand{\guard}{\phantom{m( A_{BL} ) < m( A )}}

\renewcommand{\partitionings}{
$
\phantom{\FlaTwoByTwo{A_{TL}}{A_{TR}}{A_{BL}}{A_{BR}}
\rightarrow
\FlaThreeByThreeBR
   {A_{00}}{a_{01}}{A_{02}}
   {a_{10}^T}{\alpha_{11}}{a_{12}^T}
   {A_{20}}{a_{21}}{A_{22}}}   
$
}

\renewcommand{\partitionsizes}{$ \phantom{m( A )} $}

\renewcommand{\repartitionings}{
$
\phantom{\FlaTwoByTwo{A_{TL}}{A_{TR}}{A_{BL}}{A_{BR}}
\rightarrow
\FlaThreeByThreeBR
   {A_{00}}{a_{01}}{A_{02}}
   {a_{10}^T}{\alpha_{11}}{a_{12}^T}
   {A_{20}}{a_{21}}{A_{22}}}   
$
}

\renewcommand{\repartitionsizes}{$\phantom{m(A)}$}

\renewcommand{\moveboundaries}{
$
\phantom{\FlaTwoByTwo{A_{TL}}{A_{TR}}{A_{BL}}{A_{BR}}
\rightarrow
\FlaThreeByThreeBR
   {A_{00}}{a_{01}}{A_{02}}
   {a_{10}^T}{\alpha_{11}}{a_{12}^T}
   {A_{20}}{a_{21}}{A_{22}}}   
$
}

\renewcommand{\beforeupdate}{
\phantom{\FlaTwoByTwo{A_{TL}}{A_{TR}}{A_{BL}}{A_{BR}}
\rightarrow
\FlaThreeByThreeBR
   {A_{00}}{a_{01}}{A_{02}}
   {a_{10}^T}{\alpha_{11}}{a_{12}^T}
   {A_{20}}{a_{21}}{A_{22}}}   
}

\renewcommand{\afterupdate}{
\phantom{\FlaTwoByTwo{A_{TL}}{A_{TR}}{A_{BL}}{A_{BR}}
\rightarrow
\FlaThreeByThreeBR
   {A_{00}}{a_{01}}{A_{02}}
   {a_{10}^T}{\alpha_{11}}{a_{12}^T}
   {A_{20}}{a_{21}}{A_{22}}}   
}

\renewcommand{\update}{
\phantom{$
\begin{array}{l}
\\
\\
\\
\end{array}
$}
}
}

\newcommand{\NoShow}[1]{}

\newcommand{\FlaAlgorithm}{
\begin{tabular}{| p{0.92\textwidth}|} \hline
$\mbox{\color{blue}Algorithm:~}\routinename$
\\ \whline
\partitionings \\
$\mbox{\color{blue} ~~~where~}$ \partitionsizes 
\\ 
$\mbox{\color{blue}while~} \guard \mbox{~\color{blue} do}$
\\
\ifthenelse{\equal{\blocksize}{1}}{\\}%
{%
\ifthenelse{ \equal{\blocksize}{blank} }{}%
{~~~~{\bf Determine block size $ \blocksize $}\\}%
}
~~~~ 
\repartitionings \\
~~~$\mbox{\color{blue} ~~~where~}$ \repartitionsizes
\\ \hline
~~~~  \update 
\\ \hline
~~~~ 
\moveboundaries 
\\
$\mbox{\color{blue} endwhile} $
\\ \hline 
\end{tabular}
}

\newcommand{\FlaWorksheet}{
\begin{tabular}{| c | p{0.98\textwidth} |}\hline
Step & $\mbox{\color{blue}Algorithm:~}\routinename$
\\ \hline
\rowcolor{yellow!75}
1a & $ \precondition $ 
\\ \whline
4 & 
\begin{minipage}[t]{0.9\textwidth}%
\partitionings~ \\
$\mbox{\color{blue} ~~~where~}$ \partitionsizes
\end{minipage}
\\ \hline
\rowcolor{yellow!75}
2 & $ \invariant $ 
\\ \hline
3 &$\mbox{\color{blue}while~} \guard \mbox{~\color{blue} do}$
\\ \hline 
\rowcolor{yellow!75}
2,3 & ~~~~ $ \invariant \wedge \guard$ 
\\ \hline
5a & ~~~~ \begin{minipage}[t]{0.85\textwidth}%
\ifthenelse{\equal{\blocksize}{1}}{}%
{%
\ifthenelse{ \equal{\blocksize}{blank} }{}%
{{\bf Determine block size $ \blocksize $}\\}%
}
\repartitionings~ \\
$\mbox{\color{blue} ~~~where~}$ \repartitionsizes
\end{minipage}
\\ \hline
\rowcolor{yellow!75}
6 & ~~~~ $\beforeupdate $
\\ \hline
8 & ~~~~  \update 
\\ \hline
5b & ~~~~ \begin{minipage}[t]{0.85\textwidth}%
\moveboundaries~
\end{minipage}
\\ \hline
\rowcolor{yellow!75}
7 & ~~~~ $\afterupdate $
\\ \hline
\rowcolor{yellow!75}
2 & ~~~~ $ \invariant  $ 
\\ \hline
 &$\mbox{\color{blue} endwhile} $
\\ \hline \whline
\rowcolor{yellow!75}
2,3 & $ \invariant \wedge \neg( \guard )$ 
\\ \hline
\rowcolor{yellow!75}
1b & $ \postcondition $ 
\\ \hline
\end{tabular}
}

\newcommand{\FlaWorksheetNine}{
\begin{tabular}{| c | p{0.98\textwidth} |}\hline
Step & $\mbox{\color{blue}Algorithm:~}\routinename$
\\ \hline
\rowcolor{yellow!75}
\phantom{1a} & $ \phantom\precondition $ 
\\ \whline
\phantom{4} & 
\begin{minipage}[t]{0.9\textwidth}%
\partitionings~ \\
$\mbox{\color{blue} ~~~where~}$ \partitionsizes
\end{minipage}
\\ \hline
\rowcolor{yellow!75}
\phantom{2} & $ \phantom\invariant $ 
\\ \hline
\phantom{3} &$\mbox{\color{blue}while~} \guard \mbox{~\color{blue} do}$
\\ \hline 
\rowcolor{yellow!75}
\phantom{2,3} & ~~~~ $ \phantom\invariant \phantom \wedge \phantom
                \guard $ 
\\ \hline
\phantom{5a} & ~~~~ \begin{minipage}[t]{0.85\textwidth}%
\ifthenelse{\equal{\blocksize}{1}}{}%
{%
\ifthenelse{ \equal{\blocksize}{blank} }{}%
{{\bf Determine block size $ \blocksize $}\\}%
}
\repartitionings~ \\
$\mbox{\color{blue} ~~~where~}$ \repartitionsizes
\end{minipage}
\\ \hline
\rowcolor{yellow!75}
\phantom{6} & ~~~~ $\phantom\beforeupdate $
\\ \hline
\phantom{8} & ~~~~  \update 
\\ \hline
\phantom{5b} & ~~~~ \begin{minipage}[t]{0.85\textwidth}% 
\moveboundaries~
\end{minipage}
\\ \hline
\rowcolor{yellow!75}
\phantom{7} & ~~~~ $\phantom\afterupdate $
\\ \hline
\rowcolor{yellow!75}
\phantom{2} & ~~~~ $ \phantom\invariant  $ 
\\ \hline
 &$\mbox{\color{blue} endwhile} $
\\ \hline \whline
\rowcolor{yellow!75}
\phantom{2,3} & $ \phantom\invariant \wedge \neg( \phantom\guard )$ 
\\ \hline
\rowcolor{yellow!75}
\phantom{1b} & $ \phantom\postcondition $ 
\\ \hline
\end{tabular}
}

\newcommand{\FlaWorksheetEight}{
\begin{tabular}{| c | p{0.98\textwidth} |}\hline
Step & $\mbox{\color{blue}Algorithm:~}\routinename$
\\ \hline
\rowcolor{yellow!75}
1a & $ \precondition $ 
\\ \whline
4 & 
\begin{minipage}[t]{0.9\textwidth}%
\partitionings~ \\
$\mbox{\color{blue} ~~~where~}$ \partitionsizes
\end{minipage}
\\ \hline
\rowcolor{yellow!75}
2 & $ \invariant $ 
\\ \hline
3 &$\mbox{\color{blue}while~} \guard \mbox{~\color{blue} do}$
\\ \hline 
\rowcolor{yellow!75}
2,3 & ~~~~ $ \invariant \wedge \guard $ 
\\ \hline
5a & ~~~~ \begin{minipage}[t]{0.85\textwidth}%
\ifthenelse{\equal{\blocksize}{1}}{}%
{%
\ifthenelse{ \equal{\blocksize}{blank} }{}%
{{\bf Determine block size $ \blocksize $}\\}%
}
\repartitionings~ \\
$\mbox{\color{blue} ~~~where~}$ \repartitionsizes
\end{minipage}
\\ \hline
\rowcolor{yellow!75}
6 & ~~~~ $\beforeupdate $
\\ \hline
\rowcolor{orange!50}    
8 & ~~~~  \update 
\\ \hline
5b & ~~~~ \begin{minipage}[t]{0.85\textwidth}%
\moveboundaries~
\end{minipage}
\\ \hline
\rowcolor{yellow!75}
7 & ~~~~ $\afterupdate $
\\ \hline
\rowcolor{yellow!75}
2 & ~~~~ $ \invariant  $ 
\\ \hline
 &$\mbox{\color{blue} endwhile} $
\\ \hline \whline
\rowcolor{yellow!75}
2,3 & $ \invariant \wedge \neg( \guard )$ 
\\ \hline
\rowcolor{yellow!75}
1b & $ \postcondition $ 
\\ \hline
\end{tabular}
}

\newcommand{\FlaWorksheetSeven}{
\begin{tabular}{| c | p{0.98\textwidth} |}\hline
Step & $\mbox{\color{blue}Algorithm:~}\routinename$
\\ \hline
\rowcolor{yellow!75}
1a & $ \precondition $ 
\\ \whline
4 & 
\begin{minipage}[t]{0.9\textwidth}%
\partitionings~ \\
$\mbox{\color{blue} ~~~where~}$ \partitionsizes
\end{minipage}
\\ \hline
\rowcolor{yellow!75}
2 & $ \invariant $ 
\\ \hline
3 &$\mbox{\color{blue}while~} \guard \mbox{~\color{blue} do}$
\\ \hline 
\rowcolor{yellow!75}
2,3 & ~~~~ $ \invariant \wedge \guard$ 
\\ \hline
5a & ~~~~ \begin{minipage}[t]{0.85\textwidth}%
\ifthenelse{\equal{\blocksize}{1}}{}%
{%
\ifthenelse{ \equal{\blocksize}{blank} }{}%
{{\bf Determine block size $ \blocksize $}\\}%
}
\repartitionings~ \\
$\mbox{\color{blue} ~~~where~}$ \repartitionsizes
\end{minipage}
\\ \hline
\rowcolor{yellow!75}
6 & ~~~~ $\beforeupdate $
\\ \hline
8 & ~~~~  \phantom\update 
\\ \hline
5b & ~~~~ \begin{minipage}[t]{0.85\textwidth}%
\moveboundaries~
\end{minipage}
\\ \hline
\rowcolor{orange!50}    
7 & ~~~~ $\afterupdate $
\\ \hline
\rowcolor{yellow!75}
2 & ~~~~ $ \invariant  $ 
\\ \hline
 &$\mbox{\color{blue} endwhile} $
\\ \hline \whline
\rowcolor{yellow!75}
2 & $ \invariant \wedge \neg( \guard )$ 
\\ \hline
\rowcolor{yellow!75}
1b & $ \postcondition $ 
\\ \hline
\end{tabular}
}

\newcommand{\FlaWorksheetSix}{
\begin{tabular}{| c | p{0.98\textwidth} |}\hline
Step & $\mbox{\color{blue}Algorithm:~}\routinename$
\\ \hline
\rowcolor{yellow!75}
1a & $ \precondition $ 
\\ \whline
4 & 
\begin{minipage}[t]{0.9\textwidth}%
\partitionings~ \\
$\mbox{\color{blue} ~~~where~}$ \partitionsizes
\end{minipage}
\\ \hline
\rowcolor{yellow!75}
2 & $ \invariant $ 
\\ \hline
3 &$\mbox{\color{blue}while~} \guard \mbox{~\color{blue} do}$
\\ \hline 
\rowcolor{yellow!75}
2,3 & ~~~~ $ \invariant \wedge \guard $ 
\\ \hline
5a & ~~~~ \begin{minipage}[t]{0.85\textwidth}%
\ifthenelse{\equal{\blocksize}{1}}{}%
{%
\ifthenelse{ \equal{\blocksize}{blank} }{}%
{{\bf Determine block size $ \blocksize $}\\}%
}
\repartitionings~ \\
$\mbox{\color{blue} ~~~where~}$ \repartitionsizes
\end{minipage}
\\ \hline
\rowcolor{orange!50}   
6 & ~~~~ $\beforeupdate $
\\ \hline
8 & ~~~~  \phantom\update 
\\ \hline
5b & ~~~~ \begin{minipage}[t]{0.85\textwidth}%
\moveboundaries~
\end{minipage}
\\ \hline
\rowcolor{yellow!75}
7 & ~~~~ $\phantom\afterupdate $
\\ \hline
\rowcolor{yellow!75}
2 & ~~~~ $ \invariant  $ 
\\ \hline
 &$\mbox{\color{blue} endwhile} $
\\ \hline \whline
\rowcolor{yellow!75}
2,3 & $ \invariant \wedge \neg( \guard )$ 
\\ \hline
\rowcolor{yellow!75}
1b & $ \postcondition $ 
\\ \hline
\end{tabular}
}

\newcommand{\FlaWorksheetFive}{
\begin{tabular}{| c | p{0.98\textwidth} |}\hline
Step & $\mbox{\color{blue}Algorithm:~}\routinename$
\\ \hline
\rowcolor{yellow!75}
1a & $ \precondition $ 
\\ \whline
4 & 
\begin{minipage}[t]{0.9\textwidth}%
\partitionings~ \\
$\mbox{\color{blue} ~~~where~}$ \partitionsizes
\end{minipage}
\\ \hline
\rowcolor{yellow!75}
2 & $ \invariant $ 
\\ \hline
3 &$\mbox{\color{blue}while~} \guard \mbox{~\color{blue} do}$
\\ \hline 
\rowcolor{yellow!75}
2,3 & ~~~~ $ \invariant \wedge \guard $ 
\\ \hline
\rowcolor{orange!50}   
5a & ~~~~ \begin{minipage}[t]{0.85\textwidth}%
\ifthenelse{\equal{\blocksize}{1}}{}%
{%
\ifthenelse{ \equal{\blocksize}{blank} }{}%
{{\bf Determine block size $ \blocksize $}\\}%
}
\repartitionings~ \\
$\mbox{\color{blue} ~~~where~}$ \repartitionsizes
\end{minipage}
\\ \hline
\rowcolor{yellow!75}
6 & ~~~~ $\phantom\beforeupdate $
\\ \hline
8 & ~~~~  \phantom\update 
\\ \hline
\rowcolor{orange!50}   
5b & ~~~~ \begin{minipage}[t]{0.85\textwidth}%
\moveboundaries~
\end{minipage}
\\ \hline
\rowcolor{yellow!75}
7 & ~~~~ $\phantom\afterupdate $
\\ \hline
\rowcolor{yellow!75}
2 & ~~~~ $ \invariant  $ 
\\ \hline
 &$\mbox{\color{blue} endwhile} $
\\ \hline \whline
\rowcolor{yellow!75}
2,3 & $ \invariant \wedge \neg( \guard )$ 
\\ \hline
\rowcolor{yellow!75}
1b & $ \postcondition $ 
\\ \hline
\end{tabular}
}

\newcommand{\FlaWorksheetFour}{
\begin{tabular}{| c | p{0.98\textwidth} |}\hline
Step & $\mbox{\color{blue}Algorithm:~}\routinename$
\\ \hline
\rowcolor{yellow!75}
1a & $ \precondition $ 
\\ \whline
\rowcolor{orange!50}   
4 & 
\begin{minipage}[t]{0.9\textwidth}%
\partitionings~ \\
$\mbox{\color{blue} ~~~where~}$ \partitionsizes
\end{minipage}
\\ \hline
\rowcolor{yellow!75}
2 & $ \invariant $ 
\\ \hline
3 &$\mbox{\color{blue}while~} \guard \mbox{~\color{blue} do}$
\\ \hline 
\rowcolor{yellow!75}
2,3 & ~~~~ $ \invariant \wedge \guard $ 
\\ \hline
5a & ~~~~ \begin{minipage}[t]{0.85\textwidth}%
\ifthenelse{\equal{\blocksize}{1}}{}%
{%
\ifthenelse{ \equal{\blocksize}{blank} }{}%
{{\bf Determine block size $ \phantom\blocksize $}\\}%
}
$\mbox{\phantom\repartitionings}$~ \\
$\mbox{\color{blue} ~~~where~}$ \phantom\repartitionsizes
\end{minipage}
\\ \hline
\rowcolor{yellow!75}
6 & ~~~~ $\phantom\beforeupdate $
\\ \hline
8 & ~~~~  \phantom\update 
\\ \hline
5b & ~~~~ \begin{minipage}[t]{0.85\textwidth}%
\phantom\moveboundaries~
\end{minipage}
\\ \hline
\rowcolor{yellow!75}
7 & ~~~~ $\phantom\afterupdate $
\\ \hline
\rowcolor{yellow!75}
2 & ~~~~ $ \invariant  $ 
\\ \hline
 &$\mbox{\color{blue} endwhile} $
\\ \hline \whline
\rowcolor{yellow!75}
2,3 & $ \invariant \wedge \neg( \guard )$ 
\\ \hline
\rowcolor{yellow!75}
1b & $ \postcondition $ 
\\ \hline
\end{tabular}
}

\newcommand{\FlaWorksheetThree}{
\begin{tabular}{| c | p{0.98\textwidth} |}\hline
Step & $\mbox{\color{blue}Algorithm:~}\routinename$
\\ \hline
\rowcolor{yellow!75}
1a & $ \precondition $ 
\\ \whline
4 & 
\begin{minipage}[t]{0.9\textwidth}% 
$\mbox{\phantom{\partitionings}}$~ \\
$\mbox{\color{blue} ~~~where~}$\phantom{\partitionsizes}  
\end{minipage}
\\ \hline
\rowcolor{yellow!75}
2 & $ \invariant $ 
\\ \hline
\rowcolor{orange!50}  
3 &$\mbox{\color{blue}while~} \guard \mbox{~\color{blue} do}$
\\ \hline 
\rowcolor{orange!50}   
2,3 & ~~~~ $ \invariant \wedge \guard $ 
\\ \hline
5a & ~~~~ \begin{minipage}[t]{0.85\textwidth}%
\ifthenelse{\equal{\blocksize}{1}}{}%
{%
\ifthenelse{ \equal{\blocksize}{blank} }{}%
{{\bf Determine block size $ \phantom\blocksize $}\\}%
}
$\mbox{\phantom\repartitionings}$~ \\
$\mbox{\color{blue} ~~~where~}$ \phantom\repartitionsizes
\end{minipage}
\\ \hline
\rowcolor{yellow!75}
6 & ~~~~ $\phantom\beforeupdate $
\\ \hline
8 & ~~~~  \phantom\update 
\\ \hline
5b & ~~~~ \begin{minipage}[t]{0.85\textwidth}%
\phantom\moveboundaries~
\end{minipage}
\\ \hline
\rowcolor{yellow!75}
7 & ~~~~ $\phantom\afterupdate $
\\ \hline
\rowcolor{yellow!75}
2 & ~~~~ $ \invariant  $ 
\\ \hline
 &$\mbox{\color{blue} endwhile} $
\\ \hline \whline
\rowcolor{orange!50}   
2,3 & $ \invariant \wedge \neg( \guard )$ 
\\ \hline
\rowcolor{yellow!75}
1b & $ \postcondition $ 
\\ \hline
\end{tabular}
}

\newcommand{\FlaWorksheetTwo}{
\begin{tabular}{| c | p{0.98\textwidth} |}\hline
Step & $\mbox{\color{blue}Algorithm:~}\routinename$
\\ \hline
\rowcolor{yellow!75}
1a & $ \precondition $ 
\\ \whline
4 & 
\begin{minipage}[t]{0.9\textwidth}%
$\mbox{\phantom{\partitionings}}$~ \\
$\mbox{\color{blue} ~~~where~} $ \phantom{\partitionsizes} 
\end{minipage}
\\ \hline
\rowcolor{orange!50} 
2 & $ \invariant $ 
\\ \hline
3 &$\mbox{\color{blue}while~} \phantom\guard \mbox{~\color{blue} do}$
\\ \hline 
\rowcolor{orange!50} 
2,3 & ~~~~ $ \invariant \wedge \phantom \guard $ 
\\ \hline
5a & ~~~~ \begin{minipage}[t]{0.85\textwidth}%
\ifthenelse{\equal{\blocksize}{1}}{}%
{%
\ifthenelse{ \equal{\blocksize}{blank} }{}%
{{\bf Determine block size $ \phantom\blocksize $}\\}%
}
$\mbox{\phantom\repartitionings}$~ \\
$\mbox{\color{blue} ~~~where~}$ \phantom\repartitionsizes
\end{minipage}
\\ \hline
\rowcolor{yellow!75}
6 & ~~~~ $\phantom\beforeupdate $
\\ \hline
8 & ~~~~  \phantom\update 
\\ \hline
5b & ~~~~ \begin{minipage}[t]{0.85\textwidth}%
\phantom\moveboundaries~
\end{minipage}
\\ \hline
\rowcolor{yellow!75}
7 & ~~~~ $\phantom\afterupdate $
\\ \hline
\rowcolor{orange!50} 
2 & ~~~~ $ \invariant  $ 
\\ \hline
 &$\mbox{\color{blue} endwhile} $
\\ \hline \whline
\rowcolor{orange!50} 
2 & $ \invariant \wedge \neg( \phantom\guard )$ 
\\ \hline
\rowcolor{yellow!75}
1b & $ \postcondition $ 
\\ \hline
\end{tabular}
}

\newcommand{\FlaWorksheetOne}{
\begin{tabular}{| c | p{0.98\textwidth} |}\hline
Step & $\mbox{\color{blue}Algorithm:~}\routinename$
\\ \hline
\rowcolor{orange!50}
1a & $ \precondition $ 
\\ \whline
4 & 
\begin{minipage}[t]{0.9\textwidth}%
$\mbox{\phantom{\partitionings}}$ ~ \\
$\mbox{\color{blue} ~~~where~}$ \phantom{\partitionsizes} 
\end{minipage}
\\ \hline
\rowcolor{yellow!75}
2 & $ \phantom\invariant $ 
\\ \hline
3 &$\mbox{\color{blue}while~} \phantom\guard \mbox{~\color{blue} do}$
\\ \hline 
\rowcolor{yellow!75}
2,3 & ~~~~ $ \phantom\invariant \wedge \phantom \guard$ 
\\ \hline
5a & ~~~~ \begin{minipage}[t]{0.85\textwidth}%
\ifthenelse{\equal{\blocksize}{1}}{}%
{%
\ifthenelse{ \equal{\blocksize}{blank} }{}%
{{\bf Determine block size $ \phantom\blocksize $}\\}%
}
$\mbox{\phantom\repartitionings}$ ~ \\
$\mbox{\color{blue} ~~~where~}$ \phantom\repartitionsizes
\end{minipage}
\\ \hline
\rowcolor{yellow!75}
6 & ~~~~ $\phantom\beforeupdate $
\\ \hline
8 & ~~~~  \phantom\update 
\\ \hline
5b & ~~~~ \begin{minipage}[t]{0.85\textwidth}%
\phantom\moveboundaries~
\end{minipage}
\\ \hline
\rowcolor{yellow!75}
7 & ~~~~ $\phantom\afterupdate $
\\ \hline
\rowcolor{yellow!75}
2 & ~~~~ $ \phantom\invariant  $ 
\\ \hline
 &$\mbox{\color{blue} endwhile} $
\\ \hline \whline
\rowcolor{yellow!75}
2,3 & $ \phantom\invariant \wedge \neg( \phantom\guard )$ 
\\ \hline
\rowcolor{orange!50}
1b & $ \postcondition $ 
\\ \hline
\end{tabular}
}

\newcommand{\FlaWorksheetZero}{
\begin{tabular}{| c | p{0.98\textwidth} |}\hline
Step & $\mbox{\color{blue}Algorithm:~}\routinename$
\\ \hline
\rowcolor{yellow!75}
1a & $ \phantom\precondition $ 
\\ \whline
4 & 
\begin{minipage}[t]{0.9\textwidth}%
$\mbox{\phantom{\partitionings}}$ ~ \\
$\mbox{\color{blue} ~~~where~}$ \phantom{\partitionsizes} 
\end{minipage}
\\ \hline
\rowcolor{yellow!75}
2 & $ \phantom\invariant $ 
\\ \hline
3 &$\mbox{\color{blue}while~} \phantom\guard \mbox{~\color{blue} do}$
\\ \hline 
\rowcolor{yellow!75}
2,3 & ~~~~ $ \phantom\invariant \wedge \phantom \guard$ 
\\ \hline
5a & ~~~~ \begin{minipage}[t]{0.85\textwidth}%
\ifthenelse{\equal{\blocksize}{1}}{}%
{%
\ifthenelse{ \equal{\blocksize}{blank} }{}%
{{\bf Determine block size $ \phantom\blocksize $}\\}%
}
$\mbox{\phantom\repartitionings}$~ \\
$\mbox{\color{blue} ~~~where~}$ \phantom\repartitionsizes
\end{minipage}
\\ \hline
\rowcolor{yellow!75}
6 & ~~~~ $\phantom\beforeupdate $
\\ \hline
8 & ~~~~  \phantom\update 
\\ \hline
5b & ~~~~ \begin{minipage}[t]{0.85\textwidth}%
\phantom\moveboundaries~
\end{minipage}
\\ \hline
\rowcolor{yellow!75}
7 & ~~~~ $\phantom\afterupdate $
\\ \hline
\rowcolor{yellow!75}
2 & ~~~~ $ \phantom\invariant  $ 
\\ \hline
 &$\mbox{\color{blue} endwhile} $
\\ \hline \whline
\rowcolor{yellow!75}
2,3 & $ \phantom\invariant \wedge \neg( \phantom\guard )$ 
\\ \hline
\rowcolor{yellow!75}
1b & $ \postcondition $ 
\\ \hline
\end{tabular}
}

\newcommand{\TBTinitialize}{}
\newcommand{\FlaAlgorithmTBT}{
\begin{tabular}{|l|} \hline
$\mbox{\color{blue}Algorithm:~}\routinename$
\\ \whline
\partitionings \\
$\mbox{\color{blue} ~~~where~}$ \partitionsizes 
\\ 
\TBTinitialize\\
$\mbox{\color{blue}while~} \guard \mbox{~\color{blue} do}$
\\
\ifthenelse{\equal{\blocksize}{1}}{}%
{%
\ifthenelse{ \equal{\blocksize}{blank} }{}%
{~~~~{\bf Determine block size $ \blocksize $}\\}%
}
~~~~ 
\repartitionings \\
~~~$\mbox{\color{blue} ~~~where~}$ \repartitionsizes
\\ \hline
~~~~  \update 
\\ \hline
~~~~ 
\moveboundaries 
\\
$\mbox{\color{blue} endwhile} $
\\ \hline 
\end{tabular}
}

\newcommand{\R}{\mathbb{R}}

\newcommand{\tr}[1]{{#1}^{\mathrm{T}}}

\newcommand{\trilu}[1]{\mbox{\sc trilu}( #1 )}

\newcommand{\Rmm}{\R^{m \times m}}
\newcommand{\Rnn}{\R^{n \times n}}
\newcommand{\Rmn}{\R^{m \times n}}
\newcommand{\Rmk}{\R^{m \times k}}
\newcommand{\Rkn}{\R^{k \times n}}
\newcommand{\gemm}{{\sc gemm}\xspace}
\newcommand{\trsm}{{\sc trsm}\xspace}
\newcommand{\laswp}{{\sc laswp}\xspace}
\newcommand{\panel}{{\sc panel}\xspace}
\newcommand{\gepp}{{\sc gepp}\xspace}

\journal{arXiv.org}

\begin{document}

\begin{frontmatter}

%% Title, authors and addresses

%% use the tnoteref command within \title for footnotes;
%% use the tnotetext command for theassociated footnote;
%% use the fnref command within \author or \address for footnotes;
%% use the fntext command for theassociated footnote;
%% use the corref command within \author for corresponding author footnotes;
%% use the cortext command for theassociated footnote;
%% use the ead command for the email address,
%% and the form \ead[url] for the home page:
%% \title{Title\tnoteref{label1}}
%% \tnotetext[label1]{}
%% \author{Name\corref{cor1}\fnref{label2}}
%% \ead{email address}
%% \ead[url]{home page}
%% \fntext[label2]{}
%% \cortext[cor1]{}
%% \address{Address\fnref{label3}}
%% \fntext[label3]{}

\title{A Case for Malleable Thread-Level Linear Algebra Libraries:\\
The LU Factorization with Partial Pivoting}

%% use optional labels to link authors explicitly to addresses:
 \author[UJI]{Sandra Catal\'an}
 \ead{catalans@uji.es}
 \author[UPC]{Jos\'e R. Herrero}
 \ead{josepr@ac.upc.edu}
 \author[UJI]{Enrique S. Quintana-Ort\'{\i}}
 \ead{quintana@uji.es}
 \author[UJI]{Rafael Rodr\'{\i}guez-S\'anchez}
 \ead{rarodrig@uji.es}
 \author[UTX]{Robert van de Geijn}
 \ead{rvdg@cs.utexas.edu}
 \address[UJI]{Depto. Ingenier\'{\i}a y Ciencia de Computadores, Universidad Jaume I, Castell\'on, Spain.}
 \address[UPC]{Dept. d'Arquitectura de Computadors, Universitat Polit\`ecnica de Catalunya, Spain.}
 \address[UTX]{Department of Computer Science, Institute for Computational Engineering and Sciences, The University of Texas at Austin, USA.}

\begin{abstract}

We propose two novel techniques for overcoming load-imbalance encountered when implementing so-called look-ahead mechanisms in 
relevant dense matrix factorizations for the solution of linear systems. Both techniques target the scenario where two thread teams are created/activated during the factorization,
with each team in charge of performing an independent task/branch of execution.  
The first technique promotes {\em worker sharing} (WS) between the two tasks, allowing the threads of the task that completes first to be reallocated for use by the 
costlier task. 
The second technique allows a fast task to alert the slower task of completion, enforcing the {\em early termination} (ET) of the second task,
and a smooth transition of the factorization procedure into the next iteration.

The two mechanisms are instantiated via a new {\em malleable} thread-level implementation of the {\em Basic Linear Algebra Subprograms} (BLAS),
and their benefits are illustrated via an implementation of the LU factorization with partial pivoting enhanced with look-ahead.
Concretely, our experimental results on a six core Intel-Xeon processor
show the benefits of combining WS+ET, reporting competitive performance in comparison with a task-parallel runtime-based solution.

\end{abstract}

\begin{keyword}
%% keywords here, in the form: keyword \sep keyword

%% PACS codes here, in the form: \PACS code \sep code

%% MSC codes here, in the form: \MSC code \sep code
%% or \MSC[2008] code \sep code (2000 is the default)

\end{keyword}

\end{frontmatter}

%% \linenumbers

\section{Introduction}

\newcommand{\ta}{{\sf T}$_{{\sf A}}$\xspace}
\newcommand{\tb}{{\sf T}$_{{\sf B}}$\xspace}

%Nowadays scientific computing is widely adopted for the simulation of many real world phenomena, 
%where theory is simply not appropriate whereas the economic costs of reproducing their behavior 
%by practical experimentation renders this approach unfeasible~\cite{DongarraEA11,JosephCICMM10}.
%The mathematical formulation of the physical laws modeling many real world phenomena often requires
%advanced numerical algorithms for
%linear algebra, spectral problems, N-body methods,
%partial differential equations, searching, sorting, and optimization, among others~\cite{Heath2002,Yang2008}.
%In particular, many of the computations that are needed to solve
%current scientific, engineering and industrial applications
%can be cast in terms of a reduced number of
%a reduced number of well-known dense linear algebra (DLA) problems like, e.g., simple linear algebra operations,
%linear systems of equations, rank-revealing decompositions,
%and eigenvalue problems~\cite{Heath2002,Yang2008,GVL3}.
%Moreover, the efficiency of the solvers for these DLA solvers frequently determines
%the efficacy of the application software built on top of them to simulate  and analyze these phenomena.
In the 1970s and 80s, the scientific community recognized the
value of defining standard interfaces for dense linear algebra (DLA) operations with the introduction of
the {\em Basic Linear Algebra Subprograms} (BLAS)~\cite{blas1,blas2,blas3}. 
Ever since, the key to performance portability in this domain has been the development of highly-optimized, 
architecture-specific implementations of the BLAS, 
either by hardware vendors (e.g., Intel MKL~\cite{MKL}, AMD ACML~\cite{ACML}, 
IBM ESSL~\cite{ESSL}, and NVIDIA CUBLAS~\cite{cuBLAS}) or independent developers
(e.g., GotoBLAS~\cite{Goto:2008:AHM:1356052.1356053,Goto:2008:HPI}, OpenBLAS~\cite{OpenBLAS}, ATLAS~\cite{atlas}, and BLIS~\cite{BLIS1}).

%The {\em Linear Algebra PACKage} (LAPACK)~\cite{lapack}
%defines a standard interface for a collection of routines/functionality to tackle more complex DLA operations, such as linear systems of equations
%and eigenvalue problems.
%The routines in LAPACK are formulated as blocked algorithms that perform a sequence of calls to BLAS.
%The conventional strategy to extract BDP in the legacy implementation of LAPACK, accessible at netlib,\footnote{LAPACK version 3.6.1, June 2016. 
%Available at \url{http://www.netlib.org/lapack}.} is straight-forward, as it simply relays on a multi-threaded implementation of the BLAS.

%Task-parallelism\footnote{On multi-core architectures, TP is generally exploited via the use of multiple threads and, therefore,
%is connected to BDP. 
%We distinguish between these two by
%numbers of cores in computer architectures.
Multi-threaded instances of the BLAS for current multi-core processor architectures %and many-core accelerators
take advantage of the simple data dependencies featured by these operations 
to exploit loop/data-parallelism at the block level (hereafter referred to as {\em block-data parallelism}, or BDP).
For more complex DLA operations, like those supported by LAPACK~\cite{lapack}
and {\tt libflame}~\cite{libflameref}, exploiting {\em task-parallelism} with dependencies\footnote{We view TP 
as the type of concurrency present in an operation that can be decomposed into a collection of suboperations (tasks) connected
by a rich set of data dependencies. Compared with this, BDP is present when the operation basically consists of a number of 
independent (inherently parallel) suboperations, each acting on a disjoint block of data.}
%Task-parallelism (TP) has been recently exposed as an efficient alternative to BDP in order to leverage the increasing
(TP) is especially efficient when performed
by a runtime that semi-automatically decomposes the computation into tasks and orchestrates their 
dependency-aware scheduling~\cite{ompssweb,starpuweb,plasmaweb,flameweb}.
For the BLAS kernels though, exploiting BDP is still the preferred choice, 
because it allows tighter control on the data movements across
the memory hierarchy and avoids the overhead of a runtime that is unnecessary due to the (mostly) nonexistent data dependencies in the BLAS kernels.

Exploiting both BDP and TP, in a sort of nested parallelism can yield more efficient solutions
as the number of cores in processor architectures continues to grow. For example, consider
%an architecture that supports the simultaneous execution of $t$ threads, and 
an application composed of two independent tasks, 
\ta and \tb, both of which are inherently parallel and preceded/followed by synchronization points.
In this scenario, exploiting TP only is inefficient, because it can keep at most 2 threads busy.
To address this, we could take advantage of the BDP inside \ta and \tb via TP but, 
%this may require changes inside these two tasks, and 
given their inherent parallelism, this is likely to incur certain
overhead compared with a direct exploitation of BDP. Finally, extracting BDP only is surely possible, but it may be
less scalable than a nested TP+BDP solution that splits the threads into two teams, and puts them 
to work on \ta and \tb concurrently.

Let us consider now that the previous scenario occurs during the execution of a complex DLA operation where both tasks,
\ta and \tb, can be computed via simple calls to a multi-threaded instance of BLAS. Although the solution seems
obvious, exploiting nested TP+BDP parallelism here can still be suboptimal. In particular, 
{\em all multi-threaded instances of BLAS offer a rigid interface to control the threading execution of a routine,
which only allows one to set the number of threads that will participate before the routine is invoked. Importantly, this number cannot be changed
during its execution}. Thus, in case \ta is completed earlier, the team of threads in charge of
its execution will remain idle waiting for the completion of \tb, producing a suboptimal execution from a performance perspective.
The scenario that we have described is far from being rare in DLA. To illustrate this, we will employ 
the LU factorization with partial pivoting for the solution of linear systems~\cite{GVL3}.
%that is representative of several other matrix factorizations for DLA problems.
High-performance algorithms for this decomposition consist of a loop-body that processes the matrix from 
its top-left corner to the bottom-right one, at each iteration
computing a panel factorization and updating a trailing submatrix via calls to BLAS.
We will review this factorization as a prototypical example to make the following contributions in our paper: 
\begin{itemize}
\item {\em Malleable DLA libraries}: We introduce a malleable thread-level implementation of BLIS~\cite{BLIS1}
      that allows the number of threads that participate in the execution of a BLAS kernel to dynamically change at execution time.
\item {\em Worker Sharing (WS):} %For the the LU factorization with partial pivoting and look-ahead,
      In case the panel factorization is less expensive than the update of the trailing submatrix,
      we leverage the malleable  instance of the BLAS to improve workload balancing and performance,
      by allowing the thread team in charge of the panel factorization to be reallocated to the execution of the trailing update.
%\item We propose a FLAME formulation of the blocked right-looking (RL) algorithm for the LU factorization with partial pivoting to a look-ahead strategy~\cite{Str98}
      %that exploits nested TP+BDP.
%\item We describe how to modify the implementation of BLIS matrix multiplication (\gemm) 
      %to expand the number of threads that participate in the execution of that kernel as threads complete other tasks and become idle.
%\item We leverage our prototype version of a malleable thread-level BLIS inside 
\item {\em Early Termination (ET)}: To tackle the opposite case, where panel factorization is more expensive than the update of the trailing submatrix,
      we design an ET mechanism that allows the thread team in charge of the trailing update to communicate the alternative team of this event.
      This alert forces an ET of the panel factorization, and the advance of the factorization into the next iteration. 
\item We perform a comprehensive experimental evaluation on a
      6-core Intel Xeon E5-2603 v3 processor, using execution traces to illustrate actual benefits of our approach,
      and comparing its performance to those obtained with a runtime-based solution using OmpSs~\cite{ompssweb}.
\end{itemize}
The key to our approach is that we depart from conventional instances of BLAS to instead
view the cores/threads as a pool of computational resources that, 
upon completing the execution of a BLAS/LAPACK routine, can be tapped to participate in the
execution of another BLAS/LAPACK routine that is already in progress.
This WS supports a dynamic choice of the algorithmic block size 
as the operation progresses. Furthermore, the same idea carries
over to all other major matrix decompositions for the solution of linear systems,
such as the QR, Cholesky and $LDL^T$ factorizations~\cite{GVL3}.

\section{The BLIS Implementation of Basic Linear Algebra Kernels}
\label{sec:blis}

BLIS %({\em BLAS-like Library Instantiation Software})~\cite{BLIS1}
is a framework %for instantiating the BLAS.
%A key benefit of BLIS is that
%it is a productivity multiplier, as the framework 
that allows developers to rapidly deploy new high-performance implementations of BLAS and
BLAS-like operations on current and future architectures~\cite{BLIS1}.
A key property of the BLIS open source effort is that it exposes the internal implementation of the BLAS kernels at a
finer-grain level than OpenBLAS %~\cite{Goto:2008:AHM:1356052.1356053} 
or commercial libraries
while offering performance that is competitive with GotoBLAS, OpenBLAS, Intel MKL, and 
ATLAS~\cite{BLIS2,BLIS3}.
We start by reviewing the design principles that underlie BLIS, using the implementation of \gemm as a particular case study.

Consider the matrices $A\in \Rmk$, $B \in \Rkn$, and $C \in \Rmn$.
BLIS mimics GotoBLAS %~\cite{Goto:2008:AHM:1356052.1356053} 
to implement the \gemm  kernel%
\footnote{Actually, the kernel in the BLAS interface/BLIS implementation 
for \gemm computes $C = \alpha C + \beta op(A) \cdot op(B)$, where $\alpha,\beta$ are scalars, $op(\cdot)$ 
performs an optional transposition/Hermitian-conjugation,
and $op(A)$ is $m \times k$, $op(B)$ is $k \times n$, $C$ is $m \times n$.
For simplicity, in the description we address the case where $\alpha=\beta=1$ and the operator $op(\cdot)$ does not perform
any transformation on the input matrix.}
\begin{equation}
\label{eqn:gemm}
C \mathrel{+}= A \cdot B
\end{equation}
as three nested loops around a macro-kernel plus two packing routines
(see Loops~1--3 in \figurename~\ref{fig:gotoblas_gemm}).
The macro-kernel is then implemented in terms of two additional loops around a {\em
micro-kernel} (Loops~4 and~5 in that figure).
The loop ordering embedded in BLIS, together with the packing
routines and an appropriate choice of the  BLIS cache configuration parameters ($n_c$, $k_c$, $m_c$, $n_r$ and $m_r$),
orchestrate a regular pattern of data transfers across the levels of the memory
hierarchy, and amortize the cost of these transfers with enough computation from within the micro-kernel~\cite{BLIS1} to attain near-peak performance.
In most architectures, $m_r,n_r$ are in the range 4--16; $m_c,k_c$ are in the order of a few hundreds; and $n_c$ can be up to a 
few thousands~\cite{BLIS1,BLIS2}.

\begin{figure*}[t]
\centering
%\begin{tabular}{c}
%\begin{minipage}[c]{0.9\textwidth}
%\includegraphics[width=0.9\textwidth]{Figures/gotoblas_gemm.pdf}
%\end{minipage}\\
%\\
\begin{minipage}[c]{\textwidth}
\footnotesize
\resizebox{\linewidth}{!}{
\begin{tabular}{llll}
Loop 1 &{\bf for} $j_c$ = $0,\ldots,n-1$ {\bf in steps of} $n_c$\\
Loop 2 & \hspace{3ex}  {\bf for} $p_c$ = $0,\ldots,k-1$ {\bf in steps of} $k_c$\\
&\hspace{6ex}           \textcolor{darkblue}{$B(p_c:p_c+k_c-1,j_c:j_c+n_c-1)$} $\rightarrow \textcolor{darkblue}{B_c}$ & & // Pack into $B_c$\\
Loop 3 & \hspace{6ex}           {\bf for} $i_c$ = $0,\ldots,m-1$ {\bf in steps of} $m_c$\\
&\hspace{9ex}                     \textcolor{darkred}{$A(i_c:i_c+m_c-1,p_c:p_c+k_c-1)$} $\rightarrow \textcolor{darkred}{A_c}$ & & // Pack into $A_c$ \\
\cline{2-4}
Loop 4&\hspace{9ex} {\bf for} $j_r$ = $0,\ldots,n_c-1$ {\bf in steps of} $n_r$  & & // Macro-kernel\\
Loop 5&\hspace{12ex}   {\bf for} $i_r$ = $0,\ldots,m_c-1$ {\bf in steps of} $m_r$\\
\cline{2-3}
%Loop 6&\hspace{15ex}    {\bf for} $p_r$ = $0,\ldots,k_c-1$ {\bf in steps of} $1$ & // Micro-kernel \\
&\hspace{15ex}             \textcolor{darkgreen}{$C_c(i_r:i_r+m_r-1,j_r:j_r+n_r-1)$} & // Micro-kernel \\
%&\hspace{19ex} $:=$ ~\textcolor{darkgreen}{$C_c(i_r:i_r+m_r-1,j_r:j_r+n_r-1)$}\\
&\hspace{19ex} ~$\mathrel{+}=$     ~\textcolor{darkred}{$A_c(i_r:i_r+m_r-1,0:k_c-1)$} \\
%&\hspace{19ex} ~~~$\cdot$\!~~~~\textcolor{darkblue}{$B_c(0:k_c-1,j_r:j_r+n_r-1)$} \\
&\hspace{19ex} ~~~$\cdot$~~~~\textcolor{darkblue}{$B_c(0:k_c-1,j_r:j_r+n_r-1)$} \\
%&\hspace{15ex} {\bf endfor}\\
\cline{2-3}
&\hspace{12ex} {\bf endfor}\\
&\hspace{9ex} {\bf endfor}\\
\cline{2-4}
&\hspace{6ex} {\bf endfor}\\
&\hspace{3ex} {\bf endfor}\\
&{\bf endfor}\\
\end{tabular}
}
\end{minipage}
%\end{tabular}
\caption{High performance implementation of \gemm in BLIS. In the code, $C_c \equiv C(i_c:i_c+m_c-1,j_c:j_c+n_c-1)$
is just a notation artifact, introduced to ease the presentation of the algorithm. 
In contrast, $A_c,B_c$ correspond to actual buffers that are involved in data copies.}
\label{fig:gotoblas_gemm}
\end{figure*}

The parallelization of BLIS's \gemm for multi-threaded architectures has been analyzed
for conventional symmetric multicore processors~\cite{BLIS2}, modern many-threaded architectures~\cite{BLIS3},
and asymmetric multicore processors~\cite{Catalan2016}.
%such as the IBM PowerPC A2 (16 cores/64 threads) and the Intel Xeon Phi (60 cores/240 threads).
In all these cases, the parallel implementation exploits the BDP exposed by
the nested five-loop organization of \gemm, at one or more levels
(i.e., loops), using OpenMP or POSIX threads.

A convenient option in most single-socket systems is to parallelize either Loop~3 (indexed by $i_c$), Loop~4 (indexed by $j_r$), or a combination of both~\cite{BLIS2,BLIS3,Catalan2016}.
For example, when Loop~3 is parallelized,
each thread packs a different macro-panel $A_c$ into the L2 cache
and executes a different instance of the macro-kernel.
In contrast, when Loop~4 is parallelized,
different threads will operate on independent instances of the micro-kernel, but
access the same macro-panel $A_c$ in the L2 cache.

\begin{figure*}[tb!]
\centering %
\includegraphics[width=\textwidth]{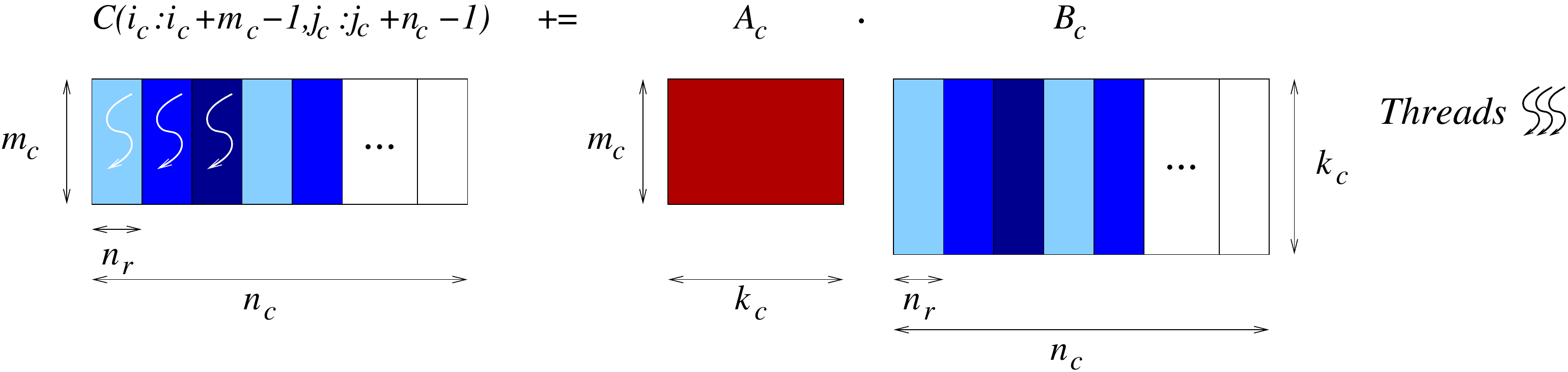}
\caption{Distribution of the workload among $t=3$~threads when Loop~4 of BLIS \gemm is parallelized. Different colors in the output
$C$ distinguish the panels of this matrix that are computed by each thread as the product of $A_c$ and corresponding panels of the input $B_c$.}
  \label{fig:parallel_loop4}
\end{figure*}

Consider, for example, a version of BLIS \gemm that extracts BDP from Loop~4 only,
to be executed on a multicore architecture with $t$ (physical) cores and one thread mapped per core.
The iteration space of Loop~4 is then statically distributed among the $t$ threads in a round-robin fashion, equivalent to 
the effect attained by adding
an OpenMP directive {\tt \#pragma omp parallel for}, with a {\tt static} schedule, 
around that loop; see Figure~\ref{fig:parallel_loop4}.
To improve performance, the packing is also performed in parallel so that,
at each iteration of Loop~3, all $t$~threads collaborate to copy and re-organize
the entries of $A(i_c:i_c+m_c-1,p_c:p_c+k_c-1)$ into the buffer $A_c$.

\section{Algorithms for the LU Factorization on Multi-threaded Architectures}
\label{sec:lu}
\newcommand{\thr}{t_{\sf ru}}
\newcommand{\thp}{t_{\sf pf}}

\newcommand{\Tp}{{\sf T}$_{{\sf PF}}$\xspace}
\newcommand{\Tr}{{\sf T}$_{{\sf RU}}$\xspace}

Given a matrix $A\in\Rmn$,
its LU factorization produces
lower and upper triangular factors, $L\in\Rmn$ and $U\in\Rnn$ respectively, such
that $PA=LU$, where $P\in\Rmm$ defines a permutation that is introduced for numerical stability~\cite{GVL3}.
In this section we first review the conventional unblocked and blocked algorithms for the LU factorization, and then
describe how BDP 
is exploited from within them. Next, we introduce a re-organized version of the algorithm that integrates look-ahead
in order to enhance performance in a nested TP+BDP execution.

All experiments in the remainder of the paper were obtained using 
{\sc ieee} double-precision arithmetic on an Intel Xeon E5-2603 v3 processor (6 
cores at a nominal frequency 1.6~GHz). The implementations were linked with BLIS version 
0.1.8 or a tailored version of this library especially developed for this work. 
Unless otherwise stated, BDP parallelism is extracted only from Loop~4 of the BLIS kernels.
%Traces were obtained using {\tt Extrae} version 3.3.0~\cite{extrae}.

\subsection{Basic algorithms and BDP}
\label{subsec:basic}

There exist a number of algorithmic variants of the LU factorization that can accommodate partial pivoting~\cite{GVL3}.
Among these, Figure~\ref{fig:LU} (left) shows an unblocked algorithm for the so-called {\em right-looking} (RL) variant,
expressed using the FLAME notation~\cite{Gunnels:2001:FFL}.
For simplicity, we do not include pivoting in the following description of the algorithms, though all our actual implementations, (and in particular
those employed in our experimental evaluation,) integrate standard partial pivoting.
The cost of computing the LU factorization of an $m\times n$ matrix, via any of the algorithms presented
in this paper, is $mn^2 - n^3/3$ floating-point
arithmetic operations (flops). Hereafter, we will consider square matrices of order $n$ for which, the cost boils down to $2n^3/3$ flops.
For the RL variants, the major part of these operations are concentrated in the
initial iterations of the algorithm(s). For example, the first 25\% iterations account for almost 58\% of the flops;
the first half for 87.5\%; and the first 75\% for more than 98\%. Thus, the key to high performance mostly lies in the initial
stages of the factorization.

\renewcommand{\blocksize}{b}

\begin{figure}[tb!]
{\footnotesize
\begin{center}
\begin{tabular}{@{} c @{} c}
\begin{minipage}{0.47\textwidth}
\setlength{\arraycolsep}{2pt}
%\footnotesize 
\renewcommand{\blocksize}{1}

\resetsteps      % Reset all the commands to create a blank worksheet  

% Define the operation to be computed

\renewcommand{\routinename}{ \left[ A \right] := \mbox{\sc LU\_unb}( A ) }

% Step 3: Loop-guard 

\renewcommand{\guard}{
  n( A_{TL} ) < n( A )
}

% Step 4: Define Initialize 

\renewcommand{\partitionings}{
  $
  A \rightarrow
  \FlaTwoByTwo{A_{TL}}{A_{TR}}
              {A_{BL}}{A_{BR}}
  $
}

\renewcommand{\partitionsizes}{
$ A_{TL} $ is $ 0 \times 0 $
}

% Step 5a: Repartition the operands 

\renewcommand{\blocksize}{1}

\renewcommand{\repartitionings}{
$  \FlaTwoByTwo{A_{TL}}{A_{TR}}
              {A_{BL}}{A_{BR}}
  \rightarrow
  \FlaThreeByThreeBR{A_{00}}{a_{01}}{A_{02}}
                    {\tr{a}_{10}}{\alpha_{11}}{\tr{a}_{12}}
                    {A_{20}}{a_{21}}{A_{22}}
$}

\renewcommand{\repartitionsizes}{
  $ \alpha_{11} $ is a scalar}

% Step 5b: Move the double lines 

\renewcommand{\moveboundaries}{
$  \FlaTwoByTwo{A_{TL}}{A_{TR}}
              {A_{BL}}{A_{BR}}
  \leftarrow
  \FlaThreeByThreeTL{A_{00}}{a_{01}}{A_{02}}
                    {\tr{a}_{10}}{\alpha_{11}}{\tr{a}_{12}}
                    {A_{20}}{a_{21}}{A_{22}}
$}

% Step 8: Insert the updates required to change the 
%         state from that given in Step 6 to that given in Step 7
% Note: The below needs editing!!!

\renewcommand{\update}{
$
  \begin{array}{cccl}
    {\sf rl1.} &
    a_{21} &:=& %l_{21} ~= 
            a_{21} / \alpha_{11}\\
%\phantom{\vspace*{20ex}}\\
~\\
~\\[0.05in]
    %\tr{a}_{12} &:=& \tr{u}_{12} = \tr{a}_{12} \\
    {\sf rl2.} &
    A_{22} &:=& A_{22} - %l_{21} \tr{u}_{12}\\
    a_{21} \tr{a}_{12}
  \end{array}
$
}

%\begin{figure}[tb!]
%\begin{center}
%\FlaAlgorithmTBT
%\end{center}
%\caption{Unblocked rigth-looking LU factorization.}
%\label{fig:LU_unb}
%\end{figure}

\scriptsize
\FlaAlgorithm
\end{minipage}
&
\begin{minipage}{0.45\textwidth}
\setlength{\arraycolsep}{2pt}
%\footnotesize 
\renewcommand{\blocksize}{b}

\resetsteps      % Reset all the commands to create a blank worksheet  

% Define the operation to be computed

\renewcommand{\routinename}{ \left[ A \right] := \mbox{\sc LU\_blk}( A ) }

% Step 3: Loop-guard 

\renewcommand{\guard}{
  n( A_{TL} ) < n( A )
}

% Step 4: Define Initialize 

\renewcommand{\partitionings}{
  $
  A \rightarrow
  \FlaTwoByTwo{A_{TL}}{A_{TR}}
              {A_{BL}}{A_{BR}}
  $
}

\renewcommand{\partitionsizes}{
$ A_{TL} $ is $ 0 \times 0 $
}

% Step 5a: Repartition the operands 

\renewcommand{\blocksize}{b}

\renewcommand{\repartitionings}{
$  \FlaTwoByTwo{A_{TL}}{A_{TR}}
              {A_{BL}}{A_{BR}}
  \rightarrow
  \FlaThreeByThreeBR{A_{00}}{A_{01}}{A_{02}}
                    {A_{10}}{A_{11}}{A_{12}}
                    {A_{20}}{A_{21}}{A_{22}}
$}

\renewcommand{\repartitionsizes}{
  $ A_{11} $ is $ b \times b $}

% Step 5b: Move the double lines 

\renewcommand{\moveboundaries}{
$  \FlaTwoByTwo{A_{TL}}{A_{TR}}
              {A_{BL}}{A_{BR}}
  \leftarrow
  \FlaThreeByThreeTL{A_{00}}{A_{01}}{A_{02}}
                    {A_{10}}{A_{11}}{A_{12}}
                    {A_{20}}{A_{21}}{A_{22}}
$}

% Step 8: Insert the updates required to change the 
%         state from that given in Step 6 to that given in Step 7
% Note: The below needs editing!!!

\renewcommand{\update}{
$
  \begin{array}{cccl}
    {\sf RL1.} & \left[ \begin{array}{c}  A_{11}\\ A_{21} \end{array}  \right] &:=& 
    %\left[ \begin{array}{c}  \{L\backslash U\}_{11}\\ L_{21}  \end{array} \right] \\
         %&=&
         \mbox{\sc LU\_unb}\left( \left[ \begin{array}{c}  A_{11}\\ A_{21}  \end{array} \right] \right)\\ [0.15in]
    {\sf RL2.} & A_{12} &:=& %U_{12} = 
                \trilu{A_{11}}^{-1} A_{12} \\
    {\sf RL3.} & A_{22} &:=& A_{22} - A_{21} A_{12}\\
  \end{array}
$
}

%\begin{figure}[tb!]
%\begin{center}
%\FlaAlgorithmTBT
%\end{center}
%\caption{Rigth-looking LU factorization with look-ahead.}
%\label{fig:Chol_blk_var3_CA}
%\end{figure}

\scriptsize
\FlaAlgorithm
\end{minipage}
\end{tabular}
\end{center}
}
\caption{Unblocked and blocked RL algorithms for the LU factorization (left and right, respectively).
         In the notation, $n(\cdot)$ returns the number of columns of its argument, and 
         $\trilu{\cdot}$ returns the strictly lower triangular part of its matrix argument, 
setting the diagonal entries of the result to ones.}
\label{fig:LU}
\end{figure}

For performance reasons, dense linear algebra libraries 
%such as LAPACK~\cite{lapack} and {\tt libflame}~\cite{libflameref},
compute the LU factorization via a blocked algorithm that casts most computations in terms of \gemm.
%in order to hide the cost of memory access on current processor architectures with hierarchical memories.
Figure~\ref{fig:LU} (right) presents the blocked RL algorithm.
For each iteration, the algorithm processes panels of $\blocksize$ columns,
where $\blocksize$ is the algorithmic block size.
The three operations in the loop body
factorize the  ``current'' panel
\(A_p = \left[ \begin{array}{c}  A_{11}\\ A_{21} \end{array}  \right]\), via the unblocked algorithm
({\sc LU\_unb}, {\sf RL1}); and next update the trailing submatrix,
consisting of
$A_{12}$ and $A_{22}$, via a triangular system solve (\trsm, {\sf RL2}) and a matrix multiplication (\gemm, {\sf RL3}), respectively.
In practice, the block size $\blocksize$ is chosen so that the successive invocations to the \gemm kernel deliver high 
FLOPS (flops per second) rates. 
If $\blocksize$ is too small, the performance of \gemm will suffer, and so will that of the
LU factorization. On the other hand, reducing $\blocksize$ is appealing as this 
choice decreases the number of flops that are performed in terms of the panel factorization, an operation that can be expected 
to offer significantly lower throughput (FLOPS) than \gemm. (Concretely, provided
$n\gg \blocksize$, the cost required for all panel factorizations is
about $n^2\blocksize/2$ flops.)  Thus, there is the tension between these two requisites.

\begin{figure*}[tb!]
\centering %
\includegraphics[width=0.6\textwidth]{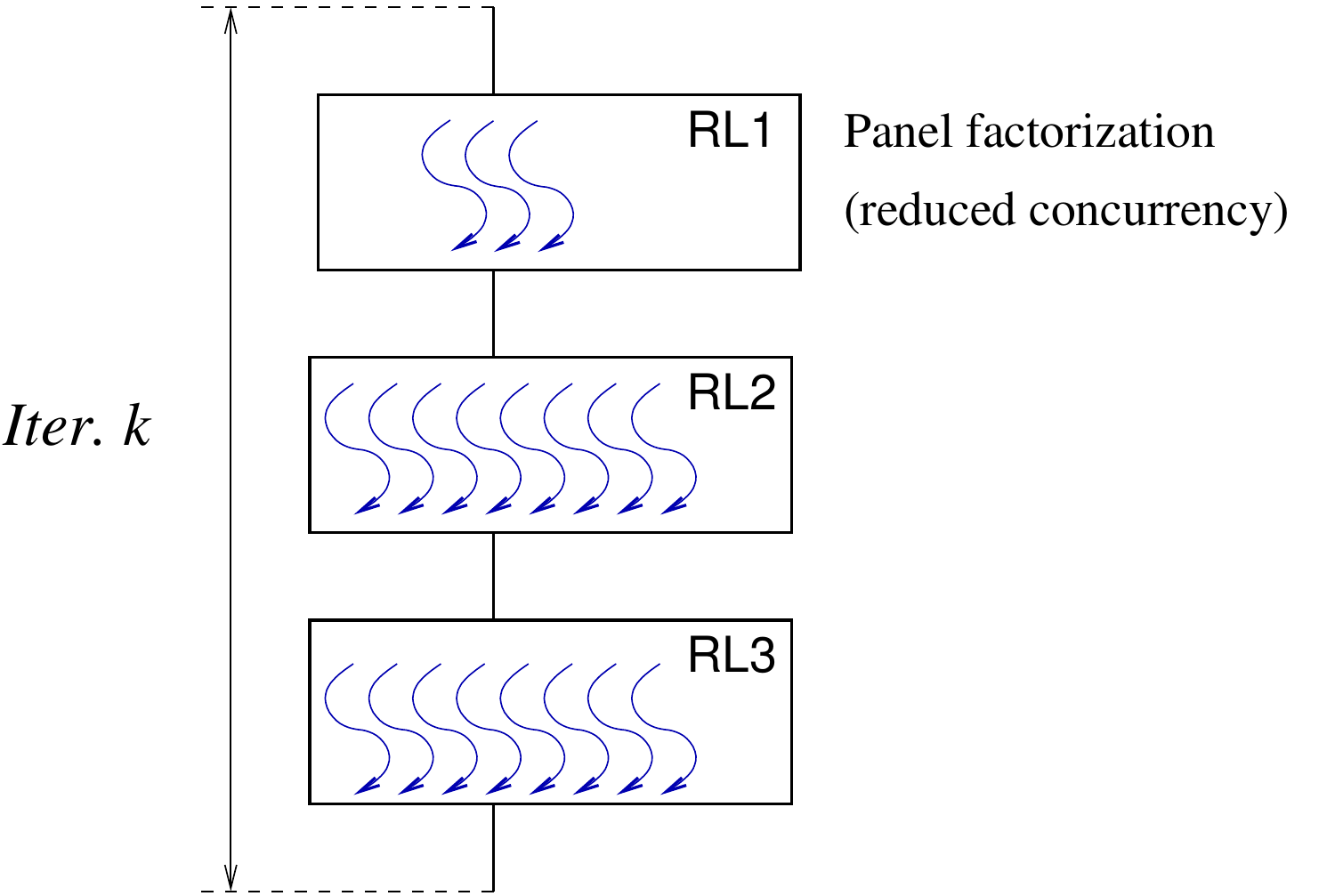}\\
\caption{Exploitation of BDP in the blocked RL LU parallelization. 
         A single thread team executes all the operations, with less active threads for {\sf RL1} due to the reduced concurrency of this kernel.
         In this algorithm, {\sf RL1} stands in the critical path.}
  \label{fig:LU_thread}
\end{figure*}

When the target platform is a multicore processor, the conventional parallelization of the LU factorization 
simply relies on multi-threaded instances of \trsm and \gemm to exploit BDP only.
%However, a particular part of the computation that stands in the critical path, known as the {\em panel factorization},
%As argued in the introduction, a straight-forward approach to parallize this algorithm is to employ a multi-threaded
%implementation of the \trsm and \gemm kernels.
Compared with this, 
the panel factorization of $A_p$, which lies in the critical path of the blocked RL factorization algorithm, 
exhibits a reduced degree of concurrency. 
Thus, depending on 
the selected block size $\blocksize$ and
certain hardware features of the target architecture (number of cores, floating-point performance, memory bandwidth, etc.),
this operation may easily become a performance bottleneck; see Figure~\ref{fig:LU_thread}. 

\begin{figure*}[tb!]
\centering %
\includegraphics[width=\textwidth]{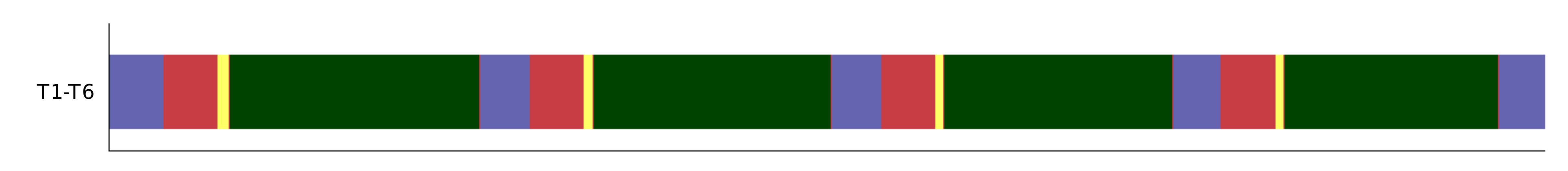}\\
\includegraphics[height=5ex]{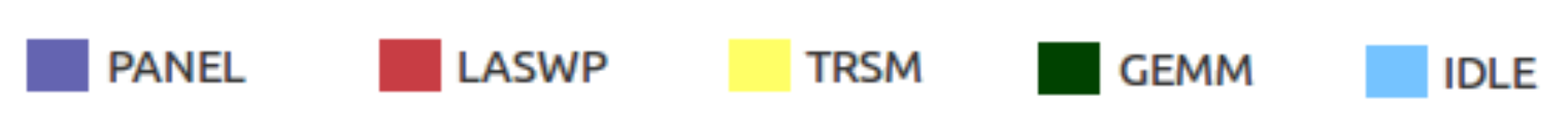}
\caption{Execution trace of the first four iterations of the blocked RL LU factorization with partial pivoting, using 6 threads, applied
         to a square matrix of order 10,000, with $\blocksize_o=256,~\blocksize_i=32$.} 
  \label{fig:trace_LU}
\end{figure*}

To illustrate the performance relevance of the panel factorization, 
Figure~\ref{fig:trace_LU} displays a fragment of a trace corresponding to the LU factorization of
a $10,000 \times 10,000$ matrix, using the blocked RL algorithm 
in Figure~\ref{fig:LU}, with partial pivoting and ``outer'' block size $\blocksize=\blocksize_o=256$. (All traces in this paper
were obtained using {\tt Extrae} version 3.3.0~\cite{extrae}.) The
code is linked with multi-threaded versions of the BLIS kernels for \gemm and \trsm.
The panel factorization (\panel) is performed via a call to the same blocked algorithm, with 
``inner'' block size $\blocksize=\blocksize_i=32$,
and also extracts BDP from the same two kernels.
With this configuration, the panel factorization represents less than 2\% of the flops
performed by the algorithm. However, the trace of the first four iterations reveals 
that its practical cost is much higher than could be expected.
(The cost of factorizing a panel relative to the cost of an iteration becomes even larger as
the iteration progresses.)
Here we note also the significant cost of the row permutations, which are performed
via the sequential legacy code for this routine in LAPACK (\laswp).
However, this second operation is embarrassingly parallel and its execution time can be expected to decrease linearly with the number of cores.

At this point, we note that the operations inside the loop body of the blocked algorithm
in Figure~\ref{fig:LU} (right) present strict dependencies that enforce their computation in the order
${\sf RL1}\Rightarrow{\sf RL2}\Rightarrow{\sf RL3}$.
Therefore, there seems to be no efficient manner to formulate a TP version of the blocked algorithm in that figure.

\subsection{Static look-ahead and nested TP+BDP}
\label{subsec:lookahead}

A strategy to tackle the hurdle represented by the panel factorization in a parallel execution
consists in the introduction of look-ahead~\cite{Str98}
into the algorithm.
Concretely, during each iteration of the decomposition 
this technique aims to overlap the factorization of the ``next'' panel with the update of the ``current'' trailing submatrix,
in practice enabling a TP version of the algorithm with two separate branches of execution, as discussed next.
%% The overlapping effect of look-ahead can be manually integrated into the code or
%implementation complexity has been considerably reduced with the
%generalization of task-parallel runtimes, which 
%% transparently attained %the same effect without requiring a re-organization of the code; see, 
%% via a task-parallel implementation supported by a run-time scheduler~\cite{BadiaHLPQQ09,Buttari200938,Quintana-Orti:2009:PMA:1527286.1527288}.
%% Unfortunately, the task-parallel approaches need to decompose the update into multiple fine-grain operations, often resulting in a suboptimal
%% use of the cache memories.
%% Look-ahead is also leveraged in Intel MKL's routine for the LU factorization%
%% \footnote{Note that we do not claim that Intel MKL employs a run-time. Since this is a proprietary (black box) library, we do not know how look-ahead is done.},
%% though this library only targets x86 architectures.

\begin{figure}[tb!]

\resetsteps      % Reset all the commands to create a blank worksheet  

% Define the operation to be computed

\renewcommand{\routinename}{ \left[ A \right] := \mbox{\sc LU\_la\_blk}( A ) }

% Step 3: Loop-guard 

\renewcommand{\guard}{
  n( A_{TL} ) < n( A )
}

% Step 4: Define Initialize 

\renewcommand{\partitionings}{
$
\begin{array}{l}
\mbox{\bf Determine block size}~b\\
  A \rightarrow
  \FlaTwoByTwo{A_{TL}}{A_{TR}}
              {A_{BL}}{A_{BR}},~
  A_{BR} \rightarrow
  \FlaOneByTwo{{\color{darkred}{A_{BR}^P}}}{{\color{darkgreen}{A_{BR}^R}}}
\end{array}
$
}

\renewcommand{\partitionsizes}{
$ A_{TL} $ is $ 0 \times 0 $, ${\color{darkred}{A_{BR}^P}}$ has $b$ columns
}

\renewcommand{\TBTinitialize}{
$ {\color{darkred}{A_{BR}^P}} := \mbox{\sc LU\_unb}\left( {\color{darkred}{A_{BR}^P}} \right)$
}

% Step 5a: Repartition the operands 

\renewcommand{\blocksize}{1}

\renewcommand{\repartitionings}{
$  \FlaTwoByTwo{A_{TL}}{A_{TR}}
              {A_{BL}}{A_{BR}}
  \rightarrow
  \FlaThreeByThreeBR{A_{00}}{A_{01}}{A_{02}}
                    {A_{10}}{A_{11}}{A_{12}}
                    {A_{20}}{A_{21}}{A_{22}}
$}

\renewcommand{\repartitionsizes}{
  $ A_{11} $ is $ b \times b $}

% Step 5b: Move the double lines 

\renewcommand{\moveboundaries}{
$  \FlaTwoByTwo{A_{TL}}{A_{TR}}
              {A_{BL}}{A_{BR}}
  \leftarrow
  \FlaThreeByThreeTL{A_{00}}{A_{01}}{A_{02}}
                    {A_{10}}{A_{11}}{A_{12}}
                    {A_{20}}{A_{21}}{A_{22}}
$}

% Step 8: Insert the updates requidarkred to change the 
%         state from that given in Step 6 to that given in Step 7
% Note: The below needs editing!!!

\renewcommand{\update}{
%\begin{array}{l}
~\\
~~~~~~{\bf Determine block size $ b $}\\
~~~~~~\% Partition into {\color{darkred}panel factorization} and {\color{darkgreen}remainder}\\
~~~~~~$
  %\begin{array}{ccl}
    %A_{12} &\rightarrow& \FlaOneByTwoSingleLine{{\color{darkred}{A_{12}^P}}}{{\color{darkgreen}{A_{12}^R}}}\\
    %A_{22} &\rightarrow& \FlaOneByTwoSingleLine{{\color{darkred}{A_{22}^P}}}{{\color{darkgreen}{A_{22}^R}}}
\renewcommand{\arraystretch}{1.2}
 \left( \begin{array}{c}  A_{12}\\\hline A_{22} \end{array}  \right) \rightarrow
 \left( \begin{array}{c|c}  {\color{darkred}{A_{12}^P}} & {\color{darkgreen}{A_{12}^R}}\\ \hline 
                           {\color{darkred}{A_{22}^P}} & {\color{darkgreen}{A_{22}^R}} \end{array}  \right) 
\renewcommand{\arraystretch}{1.0}
  %\end{array}
$\\[0.15in]
~~~~~~{\color{blue}where} both ${\color{darkred}{A_{12}^P}},~{\color{darkred}{A_{22}^P}}$ have $b$ columns\\
~\\
~~~~~~$
  \begin{array}{@{}l}
    \left. \begin{array}{@{}l}
    \mbox{\% {\color{darkred}{Panel factorization, \Tp}}}\\ [0.05in]
    {\sf PF1}.~{\color{darkred}{A_{12}^P}} := \trilu{A_{11}}^{-1} {\color{darkred}{A_{12}^P}}\\ [0.05in]
    {\sf PF2}.~{\color{darkred}{A_{22}^P}} := {\color{darkred}{A_{22}^P}} - A_{21} {\color{darkred}{A_{12}^P}}\\ [0.05in]
    {\sf PF3}.~{\color{darkred}{A_{22}^P}} := \mbox{\sc LU\_unb}\left( {\color{darkred}{A_{22}^P}} \right)
\end{array} \right. 
\hspace{0.5in}
    \left. \begin{array}{@{}l}
    \mbox{\% {\color{darkgreen}{Remainder update, \Tr}}}\\ [0.05in]
   {\sf RU1}.~{\color{darkgreen}{A_{12}^R}} := \trilu{A_{11}}^{-1} {\color{darkgreen}{A_{12}^R}}\\ [0.05in]
   {\sf RU2}.~{\color{darkgreen}{A_{22}^R}} := {\color{darkgreen}{A_{22}^R}} - A_{21} {\color{darkgreen}{A_{12}^R}}
~\\
~\\
\end{array} \right. 
  \end{array}
$

%$
    %\% Critical path 
  %\begin{array}{@{}l}
    %\left\{ \begin{array}{@{}l}
    %\mbox{\rm \% Critical path}\\
    %A_{12}^P &:=& \trilu{A_{11}}^{-1} A_{12}^P \\
    %A_{22}^P &:=& A_{22}^P - A_{21} A_{12}^P\\
    %A_{22}^P &:=& \mbox{\sc LU\_unb}\left( A_{22}^P \right)\\
%\end{array} \right. 
%\hspace{0.5in}
    %A_{33} = A_{33} - A_{31} A_{31}^T \\
    %A_{12}^R &:=& \trilu{A_{11}}^{-1} A_{12}^R \\
    %A_{22}^R &:=& A_{22}^R - A_{21} A_{12}^R\\
%$\\
%\end{array}
}

%\begin{figure}[tb!]
%\begin{center}
%\FlaAlgorithmTBT
%\end{center}
%\caption{Blocked RL algorithm enhanced with look-ahead for the LU factorization.}
%\label{fig:LU_la_blk}
%\end{figure}

%{\footnotesize
\begin{center}
\scriptsize
\FlaAlgorithmTBT
\end{center}
%}
\caption{Blocked RL algorithm enhanced with look-ahead for the LU factorization.}
\label{fig:LU_la_blk}
\end{figure}

\renewcommand{\blocksize}{b}

Figure~\ref{fig:LU_la_blk} illustrates a version of the blocked RL algorithm for the LU factorization 
re-organized to expose look-ahead.
%tackles the bottleneck due to the panel factorization by 
%overlapping the factorization of the next panel with the update of (a part of) the current trailing submatrix. 
%Consider the operations 
%{\sf RL1}, {\sf PF2} and {\sf PF3} 
%in the body loop of the standard algorithm. 
The key is to partition the trailing submatrix 
into two block column panels: 
\begin{equation}
 \left( \begin{array}{c}  A_{12}\\\hline A_{22} \end{array}  \right) \rightarrow
\renewcommand{\arraystretch}{1.2}
 \left( \begin{array}{c|c}  
         {\color{darkred}{A_{12}^P}} & {\color{darkgreen}{A_{12}^R}}\\ \hline 
         {\color{darkred}{A_{22}^P}} & {\color{darkgreen}{A_{22}^R}} \end{array}  \right) 
\renewcommand{\arraystretch}{1.0}
\label{eqn:2x2}
\end{equation}
where ${\color{darkred}{A_{22}^P}}$ corresponds to the block that, in the conventional version 
of the algorithm (i.e., without look-ahead,) would be factorized during the
next iteration. 
This effectively separates the blocks that are modified as part of the next panel factorization from the 
the remainder updates, left and right of
the $2\times 2$ partitioning in~(\ref{eqn:2x2}), respectively.
Proceeding in this manner 
creates two coarse-grain independent tasks (groups of operations in separate branches of execution): 
\Tp, consisting of {\sf PF1}, {\sf PF2}, {\sf PF3}; and
\Tr, composed of {\sf RU1} and {\sf RU2}; see Figure~\ref{fig:LU_la_blk}.
The ``decoupling'' of these block panels thus facilitate that, in a TP execution of an iteration of the loop body of the look-ahead version, the 
updates on ${\color{darkred}{A_{12}^P}},~{\color{darkred}{A_{22}^P}}$ 
and the factorization of the latter 
(operations on the next panel, in \Tp) 
proceed concurrently with the updates of
${\color{darkgreen}{A_{12}^R}},~{\color{darkgreen}{A_{22}^R}}$ 
(remainder operations, in \Tr), as there are no dependencies between \Tp and \Tr.

\begin{figure*}[tb!]
\centering %
\includegraphics[width=0.9\textwidth]{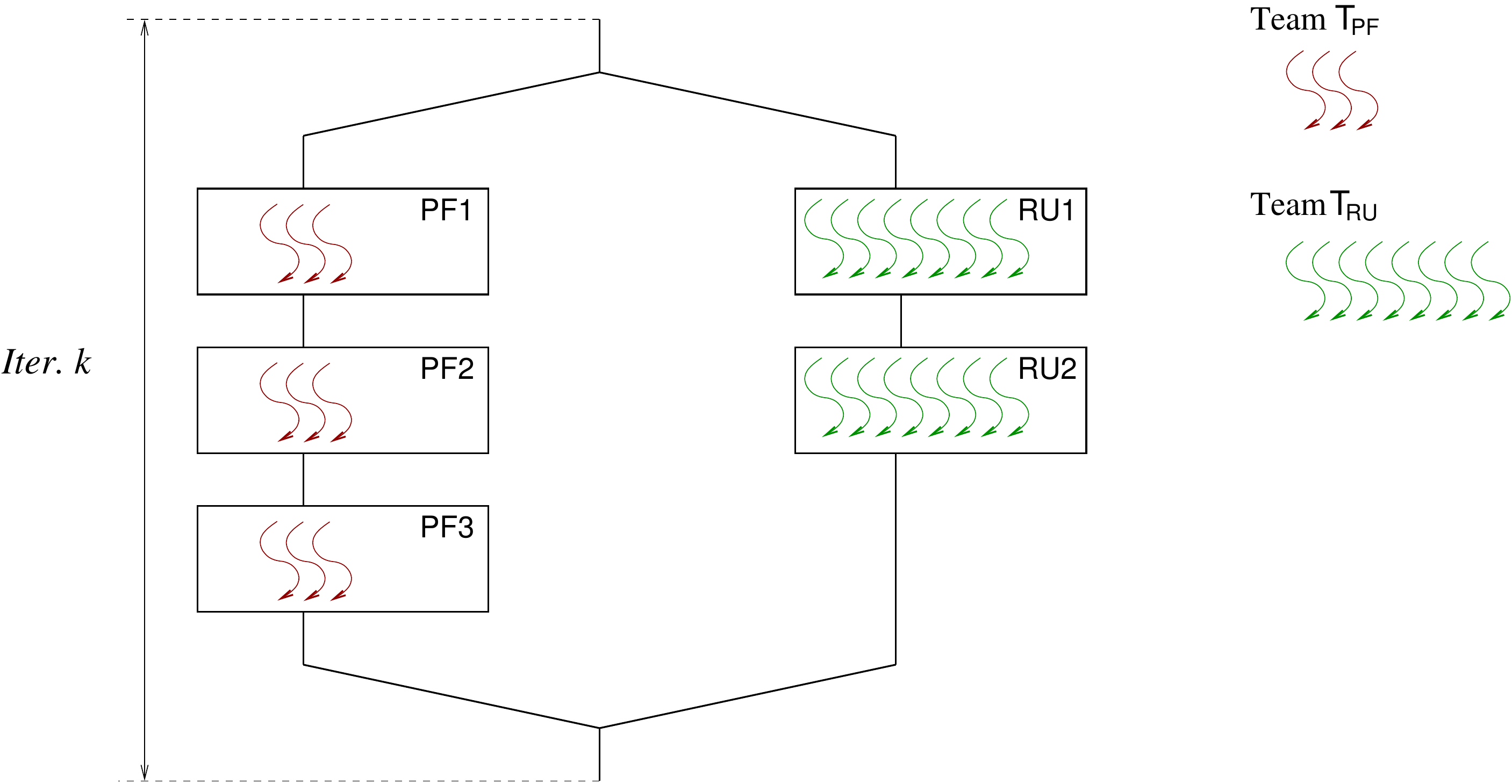}\\
\caption{Exploitation of TP+BDP in the blocked RL LU parallelization with look-ahead. 
         The execution is performed by teams \Tp and \Tr, consisting of $\thp=3$ and $\thr=8$ threads, respectively.
         In this algorithm, the operations
         on the $(k+1)$-th panel, including its factorization ({\sf PF3}), are 
         overlapped with the updates on the remainder of the trailing submatrix ({\sf RU1} and {\sf RU2}).}
  \label{fig:LU_LA_thread}
\end{figure*}

By carefully tuning the block size $\blocksize$ and adjusting the amount of computational resources (threads) dedicated to each 
of the two independent tasks, \Tp and \Tr, 
a nested TP+BDP execution of the algorithm enhanced with this static look-ahead can partially or totally overcome the bottleneck
represented by the panel factorization; see Figure~\ref{fig:LU_LA_thread}.

\begin{figure*}[tb!]
\centering %
\includegraphics[width=\textwidth]{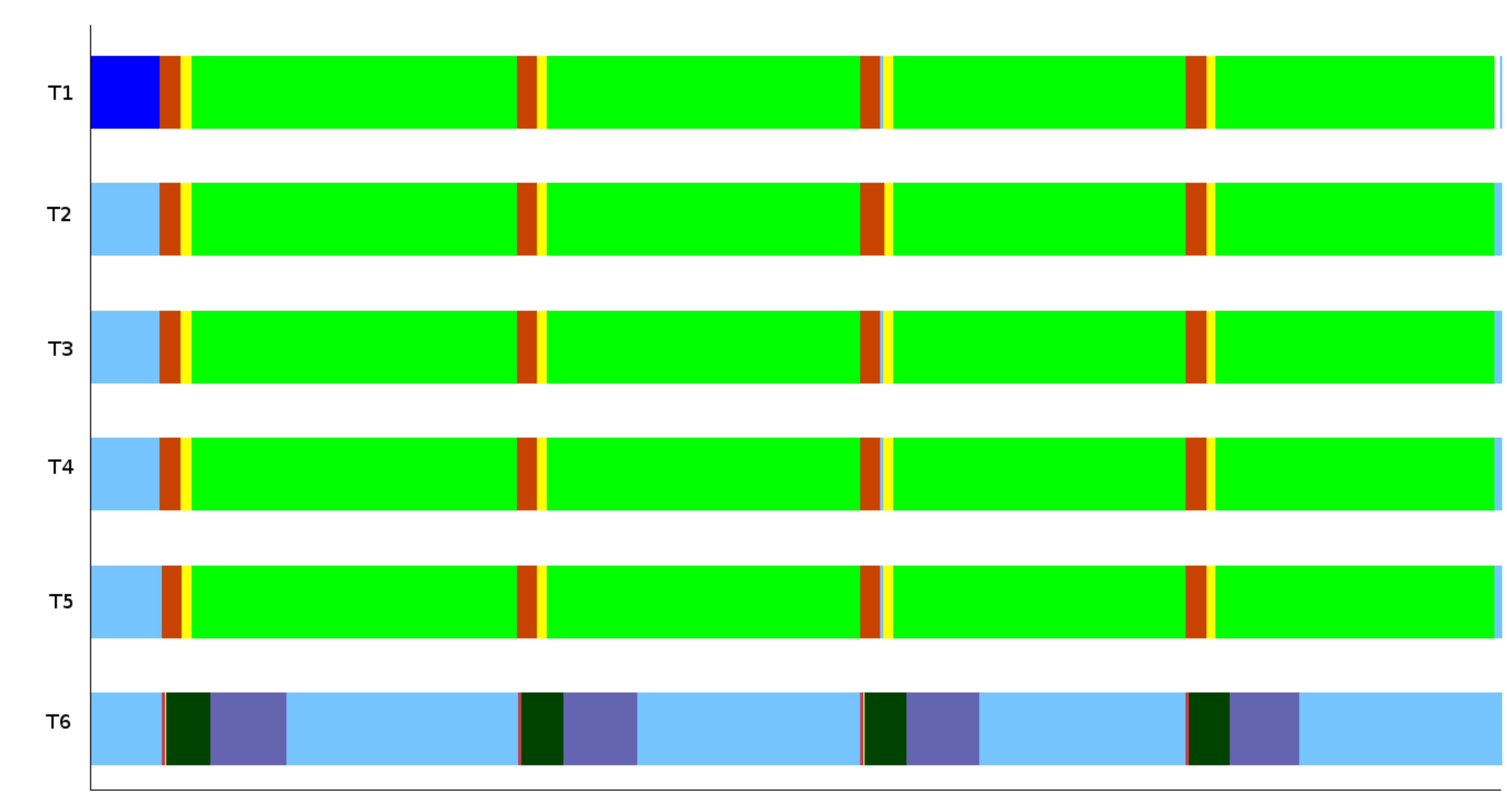}\\
\includegraphics[height=6ex,width=\textwidth]{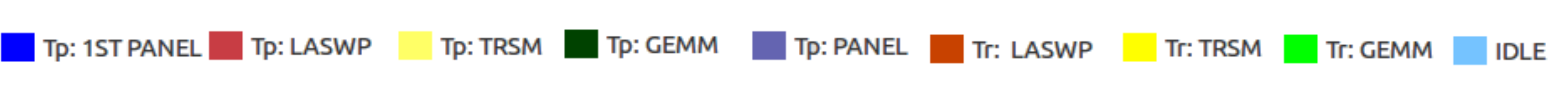}
\caption{Execution trace of the first four iterations of the blocked RL LU factorization with partial pivoting, enhanced with
         look-ahead, using 6 threads, applied to a square matrix of order 10,000, with $\blocksize_o=256,~\blocksize_i=32$.} 
  \label{fig:trace_LULA_panel}
\end{figure*}

Figure~\ref{fig:trace_LULA_panel} illustrates a complete
overlap of \Tr with \Tp attained by the look-ahead technique. The results in that figure correspond
to a fragment of a trace obtained for the LU factorization of
a $10,000 \times 10,000$ matrix, using the blocked RL algorithm 
in Figure~\ref{fig:LU_la_blk}, with partial pivoting, and outer block size $\blocksize=\blocksize_o=256$. 
For this experiment, the $t=6$~threads are partitioned into two teams: 
{\sf PF} with $\thp=1$ thread in charge of \Tp, and {\sf RU} with $\thr=5$ threads responsible for \Tr.
The panel factorization (\panel) is performed via a call to the same algorithm, with $\blocksize=\blocksize_i=32$,
and this operation proceeds sequentially (as {\sf PF} consists of a single thread).
The application of the row permutations is distributed between all 6~cores.
As argued earlier, the net effect of the look-ahead is that the cost of the panel factorization no longer has a practical impact
on the execution time of the (first four iterations of) the factorization algorithm,
which is now basically determined by the cost of the remaining operations.

\begin{figure*}[tb!]
\centering %
\includegraphics[width=\textwidth]{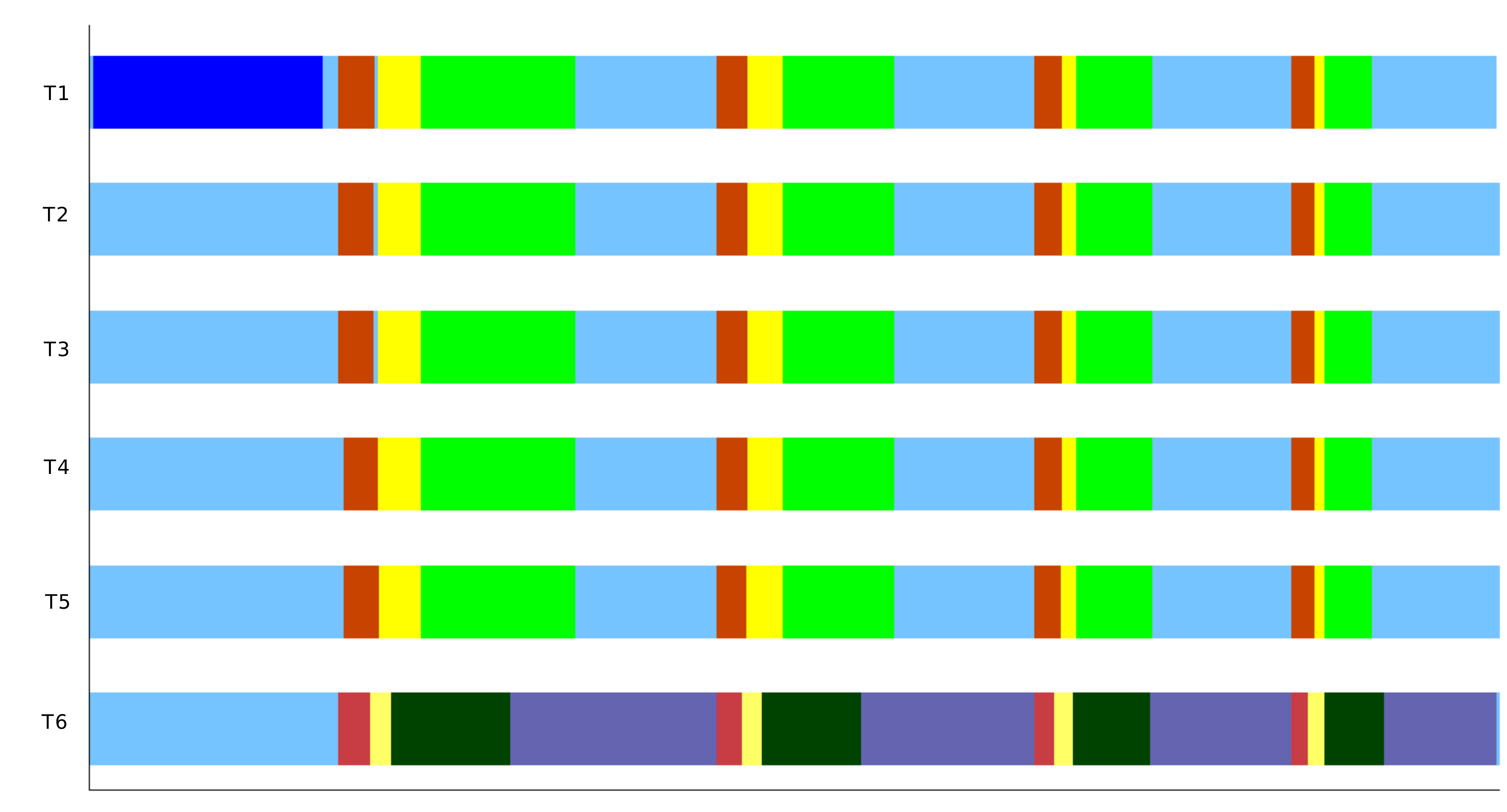}\\
\includegraphics[height=6ex,width=\textwidth]{Leyenda_resto.pdf}
\caption{Execution trace of the first four iterations of the blocked RL LU factorization with partial pivoting, enhanced with
         look-ahead, using 6 threads, applied to a square matrix of order 2,000, with $\blocksize_o=256,~\blocksize_i=32$.} 
  \label{fig:trace_LULA_update}
\end{figure*}

Given a static mapping of threads to tasks, $\blocksize$ should balance the time spent in the two tasks as,
if the operations in \Tp take longer than those in \Tr, or vice-versa, 
the threads in charge of the less expensive part will become idle, causing a performance degradation.
This was already visible in Figure~\ref{fig:trace_LULA_panel}, which shows that, during the first four iterations, the operations 
in \Tp are considerably less expensive than the updates performed as part of the remainder \Tr.
The complementary case, where \Tp requires longer than \Tr,
is illustrated using the same configuration, for a matrix of dimension $2,000\times 2,000$,
in Figure~\ref{fig:trace_LULA_update}.
Unfortunately, 
as the factorization proceeds, the theoretical costs and execution times of \Tp and \Tr vary, 
making it difficult to determine the optimal 
value of $\blocksize$, which will need to be adapted during the factorization process.

To close this section, 
note that there exist strict dependencies that serialize the operations within each task:
${\sf PF1} \Rightarrow {\sf PF2} \Rightarrow {\sf PF3}$ and
${\sf RU1} \Rightarrow {\sf RU2}$. Therefore, there is no further TP in the loop-body of this re-organized version.
However, the basic look-ahead mechanism of level/depth~1 
described in this subsection can be refined to accommodate further levels of
TP, by ``advancing'' to the current iteration the panel factorization of the following $d$ iterations, in a look-ahead
of level/depth~$d$. 
This considerably complicates the code of the algorithm, but can be seamlessly achieved by a runtime system enhanced with priorities.

\section{Advocating for Malleable Thread-Level LA Libraries}
\label{sec:lula}

For simplicity, in the following discussion we will assume that \Tp and \Tr consist only
of the panel factorization involving $A_{22}^P$ ({\sf PF3}) and the update of $A_{22}^R$ ({\sf RU2}), respectively.
%(We note that, in general, the execution times of \Tp/\Tr are indeed dominated by those of {\sf PF3}/{\sf RU2}.)
Furthermore, we will consider a nested TP+BDP execution using $t=\thp+\thr$ threads, 
{\em initially} 
with a team {\sf PF} of $\thp$ threads 
mapped to the execution of {\sf PF3} and a team {\sf RU} of $\thr$ threads computing {\sf RU2}.

%\makebox[\textwidth]{%
%\vspace*{2ex}
%\fbox{%
%\begin{minipage}{0.95\textwidth}%
Ideally, for 
the LU factorization with look-ahead, we would like to perform a {\em flexible sharing} of the computational resources so that, 
as soon as the threads in team {\sf PF} complete {\sf PF3}, they
join team {\sf RU} to help in the execution of {\sf RU2} or vice-versa.%
We next discuss these two cases in detail.
%\end{minipage}%
%} 

\subsection{Worker sharing: Panel factorization less expensive than update}

Our goal is to enable, at each iteration of the algorithm for the  LU factorization with look-ahead, 
%$\thp$ threads proceed
%with the factorization of the next panel $A_{22}^L$, while the remaining $\thr$ threads concurrently  
%perform the update of the right-hand side panel $A_{22}^R$. Eventually, 
the threads in team {\sf PF} that complete the panel factorization join the thread team {\sf RU} working
on the update. The problem is that, if the multiplication to update $A_{22}^R$ was 
initiated via an invocation to a traditional \gemm, this is not possible as
none of the existing high performance implementations of BLAS allow the number of
threads working on a kernel that is already in execution to be modified.

\subsubsection{Suboptimal solution: Static re-partitioning}
\label{subsubsec:static}

A simple workaround for this problem is to split $A_{22}^R$ into multiple column blocks, for example, 
$A_{22}^R\rightarrow\left(\begin{array}{c|c|c|c}A_1& A_2 & \ldots & A_q\end{array}\right)$,
and to perform a separate call to BLAS \gemm in order to exploit BDP during the update of each block. Then, just
before each invocation, the kernel's code
queries whether the execution of the panel factorization is completed and, if that is the case, executes the suboperation with 
the threads from both teams (or only those of {\sf RU} otherwise). Unfortunately, this approach presents several drawbacks:
\begin{itemize}
\item Replacing a single invocation to a coarse \gemm by multiple calls to smaller \gemm may offer lower throughput because the operands passed to \gemm are smaller and/or suboptimally ``shaped''. 
%This is partially due because
The consequence is that calling \gemm multiple times will internally incur re-packing and data movement overheads, which are more difficult to amortize because of the smaller problem dimensions.
\item The burden of which loop to partition for parallelism (note that $A_{22}^R$ could have alternatively been split by rows, or into blocks), 
      and the granularity of this partitioning is then placed upon the programmer's shoulders, who may lack the information
      that is necessary to make a good choice. For example, if the granularity is too coarse, this will have negative effect because the integration of the single thread
      in the update will likely be delayed. A granularity that is too fine, on the other hand, may reduce the parallelism within the BLAS operation or 
      result in the use of cache blocking parameters that are too small.
\end{itemize}

%\subsubsection{Dynamic partitioning via runtime}

%A task-parallel solution that relies on a run-time for dynamic scheduling of the 
%tasks~\cite{BadiaHLPQQ09,Buttari200938,Quintana-Orti:2009:PMA:1527286.1527288} may solve some of these problems, but not all. 
%In particular, a task-parallel approach still suffers from re-packing and data movement overheads due to multiple calls to \gemm. 
%Moreover, it couples the algorithmic block size that fixes the granularity of the tasks to that of the suboperands in \gemm.

\subsubsection{Our solution: Malleable thread-level BLAS}
\label{subsubsec:mBLAS}

The alternative that we propose in this work exploits BDP inside
{\sf RU2},
%{\sf PF2},
%{\sf PF3},
%{\sf RU1} and
%{\sf RU2} as individual tasks that can each 
{\em but allows to change the number of threads that participate in this computation even if the task is already in execution!}
In other words, we view the threads as a resource pool of workers that can be shared between different tasks and 
reassigned to the execution of a (BLAS) kernel that is already in progress.

\begin{figure*}[tb!]
\centering %
\includegraphics[width=0.9\textwidth]{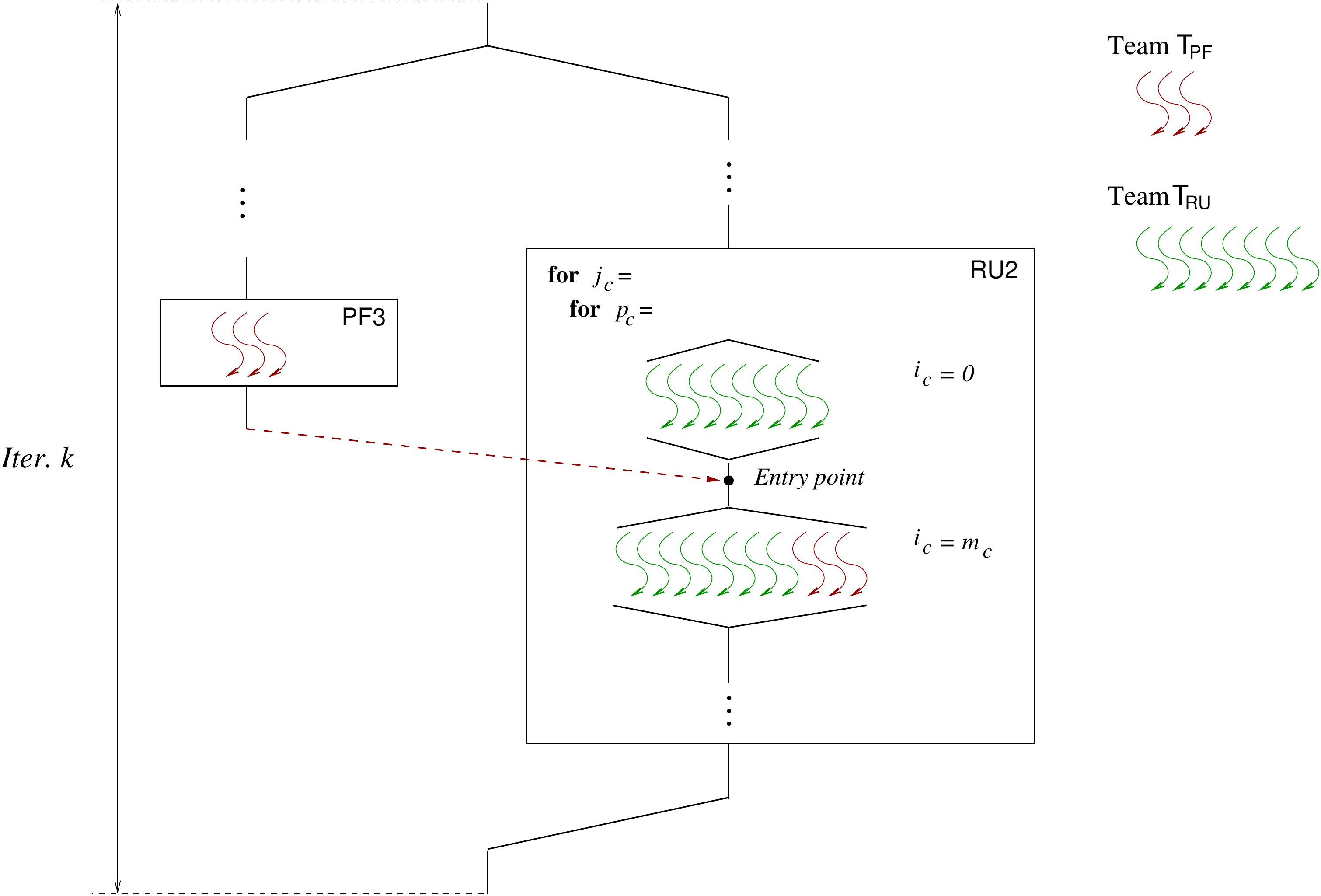}\\
\caption{Exploitation of TP+BDP in the blocked RL LU parallelization with look-ahead and WS. 
         The execution is performed by teams \Tp and \Tr, consisting of $\thp=3$ and $\thr=8$ threads, respectively.
         In this example, team \Tp completes the factorization {\sf PF3} while team \Tr is executing
         the first iteration of Loop~3 that corresponds to {\sf RU2}/\gemm ($i_c=0$). Both teams then
         merge and jointly continue the update of the remaining iterations of that loop ($i_c=m_c, 2m_c,\ldots$).
         With the parallelization of \gemm Loop~4,
         one such ``entry point'' enables the merge at the beginning of each iteration of loop $i_c$.}
  \label{fig:LU_LA_WS}
\end{figure*}

The key to our approach lies in the explicit exposure of the \gemm internals (and other BLAS-3 kernels) in BLIS. 
Concretely, assume that {\sf RU2} is computed via a single invocation to BLIS \gemm, and consider
that this operation is parallelized by distributing 
the iteration space of Loop~4 among the threads in team {\sf PF}; (see Figures~\ref{fig:gotoblas_gemm} and~\ref{fig:parallel_loop4}).
Then, just before Loop~4, %each packing of $A_c$, 
we force the system to check 
if the execution of the panel factorization is completed and, based on this information, decides whether %the packing and Loop~4 are 
this loop is executed using either the union of the threads from both teams or only those in {\sf RU};
see Figure~\ref{fig:LU_LA_WS}.
%{\bf Pending: Use figure to illustrate this}

Let us re-analyze the problems listed in Subsection~\ref{subsubsec:static} for the work-around that 
statically partitioned the update of $A_{22}^R$, and compare
them with our solution that implicitly embeds this partitioning inside BLIS:
\begin{itemize}
\item The partitioning of \gemm into multiple calls to smaller 
      matrix multiplications does not occur. Our solution performs a single call to \gemm only, so that
      there is no additional re-packing nor data movements. 
      For example, in the case just discussed, $B_c$ is already packed and re-used independently
      of whether $t$ or $\thr$ threads participate in the \gemm. The buffer
       $A_c$ is packed only once per iteration of Loop~3 (in parallel by 
      both teams or only {\sf RU}).
\item The decision of the best partitioning/granularity is left in the hands of BLIS, which likely has more information to do a better job than the programmer.
\end{itemize}
Importantly, the partitioning happens dynamically and is transparent to the programmer.

\begin{figure*}[tb!]
\centering %
\includegraphics[width=\textwidth]{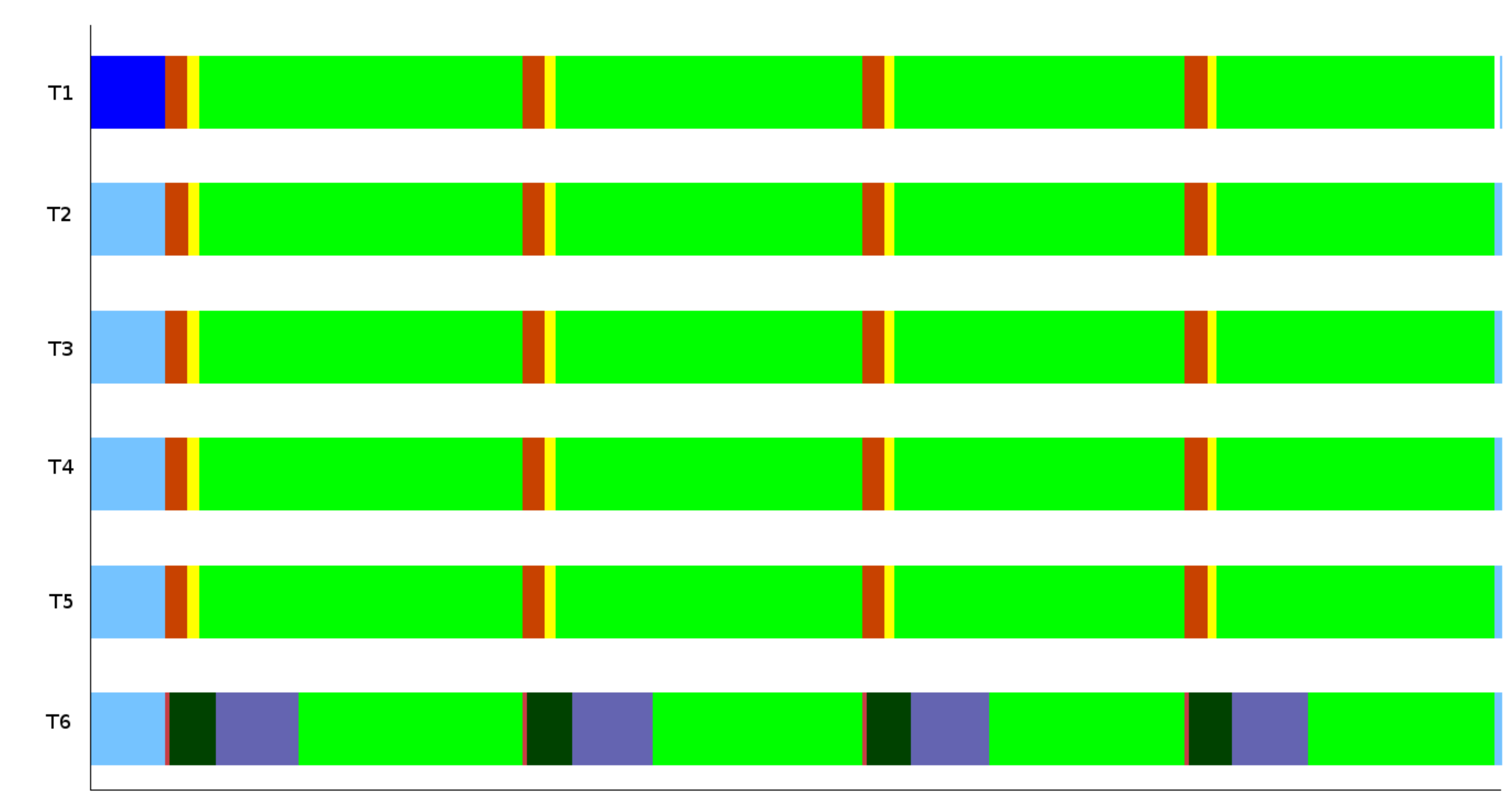}\\
\includegraphics[height=6ex,width=\textwidth]{Leyenda_resto.pdf}
\caption{Execution trace of the first four iterations of the blocked RL LU factorization with partial pivoting, enhanced with
         look-ahead and {\em malleable BLIS}, using 6 threads, applied to a square matrix of order 10,000, with $\blocksize_o=256,~\blocksize_i=32$.}
  \label{fig:trace_LULA_malleable}
\end{figure*}

Figure~\ref{fig:trace_LULA_malleable}
validates the effect of integrating a malleable version of BLIS into 
the same configuration that produced the results in Figure~\ref{fig:trace_LULA_panel}.
A comparison of both figures shows that, with a malleable version of BLIS, 
the thread executing the operations in \Tp, after completing
this task, rapidly joins the team that computes the remainder updates,
thus avoiding the idle wait.

Compared with BLIS, the same approach cannot be integrated into GotoBLAS 
because the implementation of \gemm in this library only exposes the three outermost loops of \figurename~\ref{fig:gotoblas_gemm}, while the
remaining loops are encoded in assembly.
The BLAS available as part commercial libraries is not an option either
because hardware vendors offer black-box implementations which do not permit the migration of threads.

\subsection{Early termination: Panel factorization more expensive than update}
\label{subsec:earlystop}

\begin{figure*}[tb!]
\centering %
\includegraphics[width=0.75\textwidth]{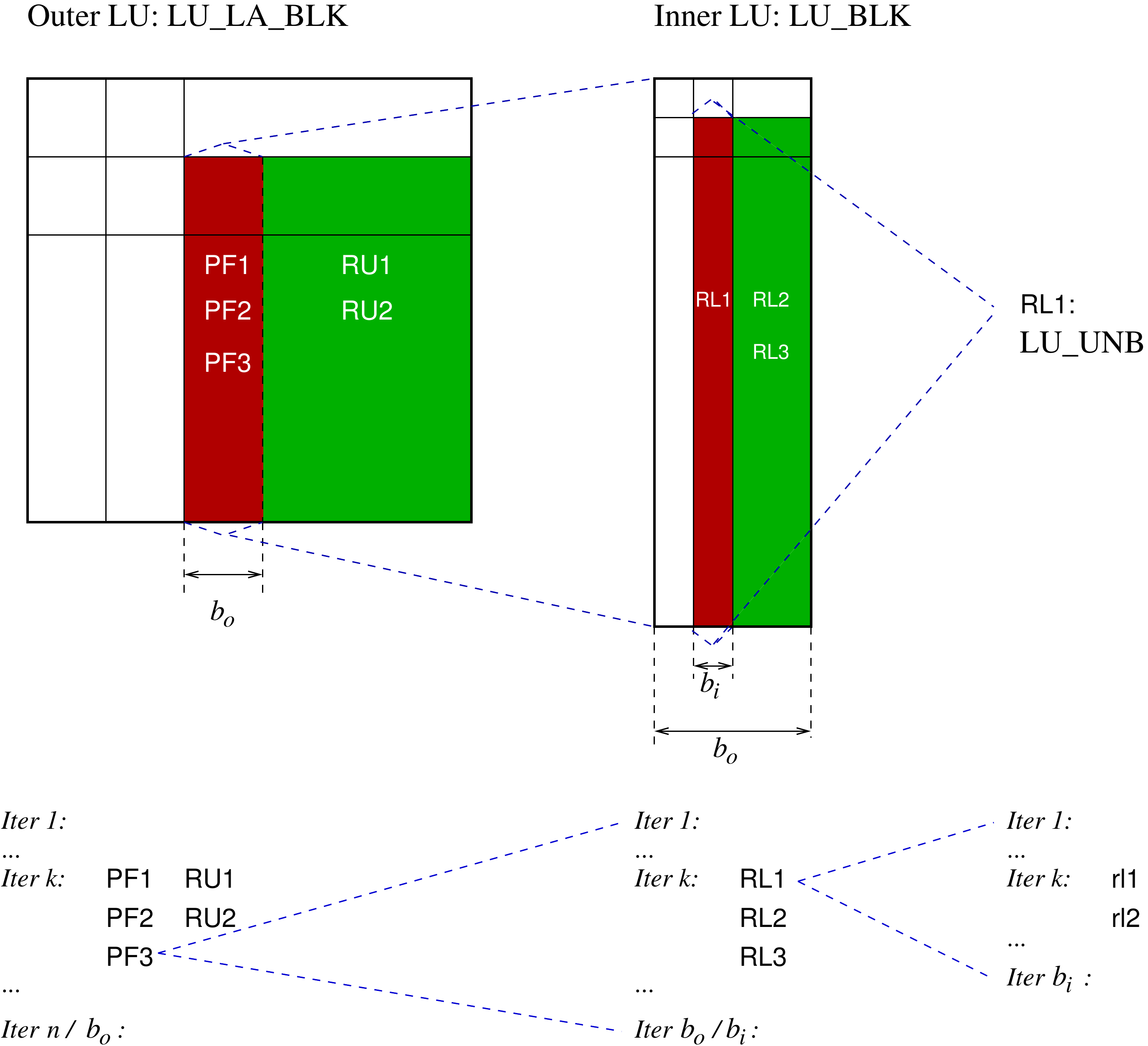}\\
\caption{Outer vs inner LU and use of algorithmic block sizes.}
  \label{fig:outer_inner_LU}
\end{figure*}

The analysis of this case will reveal some important insights.
In order to discuss them, let us consider that, in the LU factorization with look-ahead, the panel factorization 
({\sf PF3}) is performed via 
a call to the blocked routine in Figure~\ref{fig:LU} (right). %. In other words,
%{\sf PF3} in Figure~\ref{fig:LU_la_blk} becomes:
%\[
 %\begin{array}{ccl}
%{\sf PF3}.~{\color{black}{A_{22}^P}} := \mbox{\sc LU\_blk}\left( {\color{black}{A_{22}^P}} \right)
%\end{array}
%\]
%
%Here we will assume the use of two blocking
%parameters, 
We assume two blocking parameters: $\blocksize=\blocksize_o$ 
for the outer routine that computes the LU factorization of the complete matrix 
using look-ahead, and $\blocksize=\blocksize_i$ for the inner routine that factorizes each panel. % via the call in {\sf PF3}, 
%with $\blocksize_i \leq \blocksize_o$.
(Note that, if $\blocksize_i = \blocksize_o$ or $\blocksize_i$=1, 
the panel factorization is then simply done via the unblocked algorithm.)
Furthermore, we will distinguish these two levels
by referring to them as the {\em outer LU} (factorization with look-ahead) and the {\em inner LU} (factorization via the
blocked algorithm without look-ahead). 
%Thus, the outer LU operates on an (outer-)panel of width $\blocksize_o$, using a blocked algorithm with block size $\blocksize_i$; the inner LU
%operates on an inner-panel of width $\blocksize_i$, using the unblocked algorithm.
Thus, at each iteration of the outer LU, a panel of $\blocksize_o$ columns is factorized
via a call to {\sc LU\_blk} (inner LU), and this second decomposition proceeds to factorize the panel
using a blocked algorithm with block size $\blocksize_i$;
see Figure~\ref{fig:outer_inner_LU}. 

From Figure~\ref{fig:LU} (right), the loop body for the inner LU consists
of a call to the unblocked version of the algorithm ({\sf RL1}), followed by the invocations to \trsm and \gemm
that update $A_{12}$ and $A_{22}$, respectively ({\sf RL2} and {\sf RL3}). 
Now, let us assume that the update {\sf RU2} by the thread team {\sf RU} is completed 
while the threads of team {\sf PF}
are in the middle of the computations corresponding to an iteration of the loop body of the inner LU.
Then, provided
the versions of the BDP versions \trsm and \gemm kernels that are invoked from the inner LU
are malleable (see subsection~\ref{subsubsec:mBLAS}), inside them the system will perform the
actions that are necessary to integrate the thread team {\sf RU},
which is now idle, into the corresponding (and subsequent) computation(s).
Unfortunately, the updates in the loop body of the inner LU involve 
small-grained computations ($A_{12}$ and $A_{22}$ have at most $b_o-b_i$ columns, decreasing by $b_i$ columns at each iteration),
and little parallel performance can be expected from it especially because of partial pivoting.

In order to deal with this scenario, a different option is to force
%A different option is to deal with this scenario is to force
 the inner LU to stop at the end of the current iteration, to then rapidly 
proceed to the next iteration of the outer LU.
We refer to this strategy as the {\em early termination} (ET). 
In order to do this though, the transformations computed to factorize the current
inner-panel must be propagated first to the remaining columns outside this panel, introducing a certain delay in this version of the ET strategy.
%This may seem problematic as, at the end of an iteration of the inner LU, we have computed the factorization till a certain 
%point (column $k\cdot b_i$) of the panel, but we have also propagated the corresponding transformations to the remaining columns of the panel.
%A possibility is to work on a copy of the panel, and to transfer back only the first $k\cdot b_i$ columns of this workspace
%to the original matrix, thus discarding the updates to the right of the block,
%and setting $b_o=k \cdot b_i$ for this iteration.
%However, this option wastes some resources by performing certain operations which will need to be recomputed later.

A third possibility is to rely on a left-looking (LL) version of the LU factorization for the inner LU,
as discussed next. %We next elaborate this more appealing alternative further.
The blocked LL algorithm for the LU factorization differs from the blocked RL variant (see the algorithm in
the right-hand side of Figure~\ref{fig:LU})
in the operations performed inside the loop-body, which are replaced by
\[
 \begin{array}{cccl}
    {\sf LL1.} & A_{01} &:=& \trilu{A_{00}}^{-1} A_{01} \\ [0.1in]
    %{\sf LL2.} & A_{11} &:=& A_{11} - A_{10} A_{01}\\
    %{\sf LL3.} & A_{21} &:=& A_{21} - A_{20} A_{01}\\
    {\sf LL2.} & \left[ \begin{array}{c}  A_{11}\\ A_{21} \end{array}  \right] &:=& 
                 \left[ \begin{array}{c}  A_{11}\\ A_{21} \end{array}  \right] -
                 \left[ \begin{array}{c}  A_{10}\\ A_{20} \end{array}  \right] A_{01} \\ [0.2in]
    {\sf LL3.} & \left[ \begin{array}{c}  A_{11}\\ A_{21} \end{array}  \right] &:=& 
                 \mbox{\sc LU\_unb}\left( \left[ \begin{array}{c}  A_{11}\\ A_{21}  \end{array} \right] \right)
  \end{array}
\] 
Thus, at the end of a certain iteration, this variant has only updated the current column of the inner-panel and those to its left. In other words,
no transformations are propagated beyond that point (i.e., to the right of the current column/inner-panel),
and ET can be implemented in a straight-forward manner, 
with no delay compared with an inner LU factorization via the RL variant.

A definitive advantage of the LL variant compared with its RL counterpart is that the former implements a lazy algorithm, 
which delays the operations
towards the end of the panel factorization, while the second
corresponds to an eager algorithm that advances as much computations as possible to the 
initial iterations.
Therefore, in case the panel factorization has to be stopped early, it
is more likely that the LL variant has progressed in the factorization 
further.\footnote{Consider the factorization of an $m\times n$ matrix that is stopped at iteration $k<n$. The LL
algorithm will have performed $m^2k-m^3/3$ flops at that point while, for the RL algorithm,
the flop count raises to that of the LL algorithm plus $2(n-k)(mk-k^2/2)$.}
The appealing consequence is that
this enables the use of larger block sizes for the following updates in the LL variant.

\begin{figure*}[tb!]
\centering %
\includegraphics[width=0.9\textwidth]{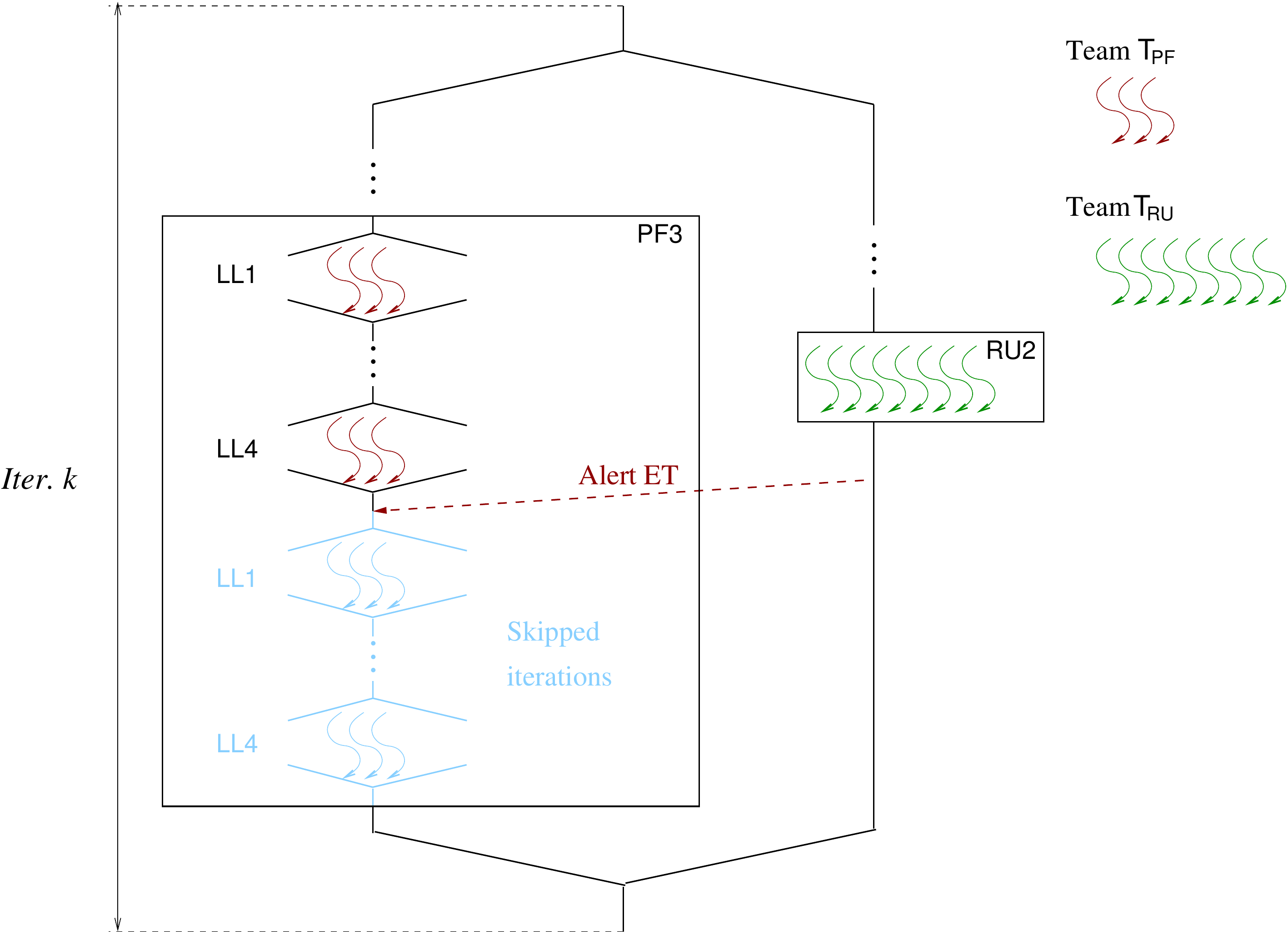}\\
\caption{Exploitation of TP+BDP in the blocked RL LU parallelization with look-ahead and ET. 
         The execution is performed by teams \Tp and \Tr, consisting of $\thp=3$ and $\thr=8$ threads, respectively.
         In this example, team \Tr completes the update {\sf RU2} while team \Tp is executing
         an iteration of the panel factorization {\sf PF3}. \Tr then notifies of this event to \Tp, which
         then skips the remaining iterations of the loop that processes the panel.}
  \label{fig:LU_LA_ET}
\end{figure*}

From an implementation point of view, the synchronization between the two teams of threads is easy to handle.
For example, at the beginning of each iteration of the outer LU, a boolean flag is set to indicate 
that the remainder update is incomplete. 
The thread team {\sf RU} then changes this value as soon as this task is complete.
In the mean time, the flag is queried by the thread team {\sf PF}, at the end of every
iteration of the inner LU, aborting its execution when a change is detected.
With this operation mode, there is no need to protect the flag from race conditions.
This solution also provides an adaptive (automatic) configuration of the block size as, if chosen too large,
it will be adjusted for the current (and, possibly, subsequent) iterations by the early termination of the inner LU.
The process is illustrated in Figure~\ref{fig:LU_LA_ET}.

%\subsection{Limitations of BLIS (and OpenMP) and partial solution}

%Discuss why we cannot implement the ideal solution with the current BLIS.

\subsection{Relation to adaptive look-ahead via a runtime}

Compared with our approach, which only applies look-ahead at one level, a TP execution that relies on a run-time for adaptive-depth look-ahead
exposes a higher degree of parallelism from ``future iterations'', which can 
amortize the cost of the panel factorization over a considerably
larger number 
of flops. This can be beneficial for architectures with a large number of cores,
but can be partially compensated by increasing the number of threads dedicated to the panel factorization, combined with a careful fine-grain exploitation
of the concurrency~\cite{Castaldo:2013:SLP:2491491.2491492}, in our approach. 
On the other hand, adaptive-depth look-ahead via a runtime suffers from re-packing and data movement overheads due to multiple calls to \gemm.
Moreover, it couples the algorithmic block size that fixes the granularity of the tasks to that of the suboperands in \gemm.
Finally, runtime-based solutions rarely exploit nested TP+BDP parallelism and, even if they do so, taking advantage of a malleable thread-level BLAS
from within them may be difficult.

\section{Experimental Evaluation}
\label{sec:experiments}
\newcommand{\bo}{\blocksize_o}

In this section we analyze in detail the performance behavior of several multi-threaded implementations of the algorithms 
for the LU factorization:
\begin{itemize}
\item {\sf LU}: Blocked RL (Figure~\ref{fig:LU}). This code only exploits BDP, via calls to the (non-malleable) multi-threaded BLIS
      (version 0.1.8).
\item Variants enhanced with look-ahead (Figure~\ref{fig:LU_la_blk}). The following three implementations
      take advantage of nested TP+BDP, with 1~thread dedicated to the operations
      on the panel (team {\sf $T_{PF}$}) and $t-1$ to the remainder updates (team {\sf $T_{RU}$}).
\begin{itemize}
\item {\sf LU\_LA} (subsection~\ref{subsec:lookahead}): Blocked RL with look-ahead.
\item {\sf LU\_MB} (subsection~\ref{subsubsec:mBLAS}): Blocked RL with look-ahead and malleable BLIS.
\item {\sf LU\_ET} (subsection~\ref{subsec:earlystop}): Blocked RL with look-ahead, malleable BLIS, and early termination of the panel factorization.
\end{itemize}
\item {\sf LU\_OS}: Blocked RL with adaptive look-ahead extracted via the OmpSs runtime (version 16.06). 
{\sf LU\_OS} decomposes the factorization into a large collection of tasks connected
via data dependencies, and then exploits TP only, via calls to a sequential instance of BLIS.
In more detail, the OmpSs parallel version divides the matrix into a collection of panels of fixed width $\bo$.
All operations performed during an iteration of the algorithm on the same panel 
(row permutation, triangular system solve, matrix multiplication and, possibly, panel factorization) are then
part of the same task. This implementation includes priorities to advance the schedule of tasks involving panel factorizations.
\end{itemize}
All codes include standard partial pivoting and compute the same factorization. 
%The flop count for this decomposition is $2/3n^3$ flops, for a square matrix of order $n$.
Also, all solutions perform the panel factorization via the blocked RL algorithm, except for
{\sf LU\_ET} and {\sf LU\_OS}, which employ the blocked LL variant. 
The performance differences between the LL and RL variants, when applied solely to the panel factorization,
were small. Nonetheless, for {\sf LU\_ET}, employing the LL variant improves the ET mechanism and unleashes a
faster execution of the global factorization. For {\sf LU\_OS} we integrated the LL variant as well to favor
a fair comparison between this implementation and our {\sf LU\_ET}.
The block size is fixed to $\bo$ during the complete iteration in all cases, 
except for 
{\sf LU\_ET} which initially employs $\bo$,  but then adjusts this value during the factorization
as part of the ET mechanism.

In the experiments, we considered the factorization of square matrices, with random
entries uniformly distributed in $(0,1)$, and dimension
$n$=500 to 12,000 in steps of 500. The block size for the outer LU was tested for
values 
$\bo$=32 to 512 in steps of 32. The block size for the inner LU was evaluated for 
$\blocksize_i$=16 and 32.
We employed one thread per core (i.e., $t=6$) in all executions.

\subsection{Optimal block size}

\begin{figure*}[tb!]
\centering %
\includegraphics[width=0.48\textwidth]{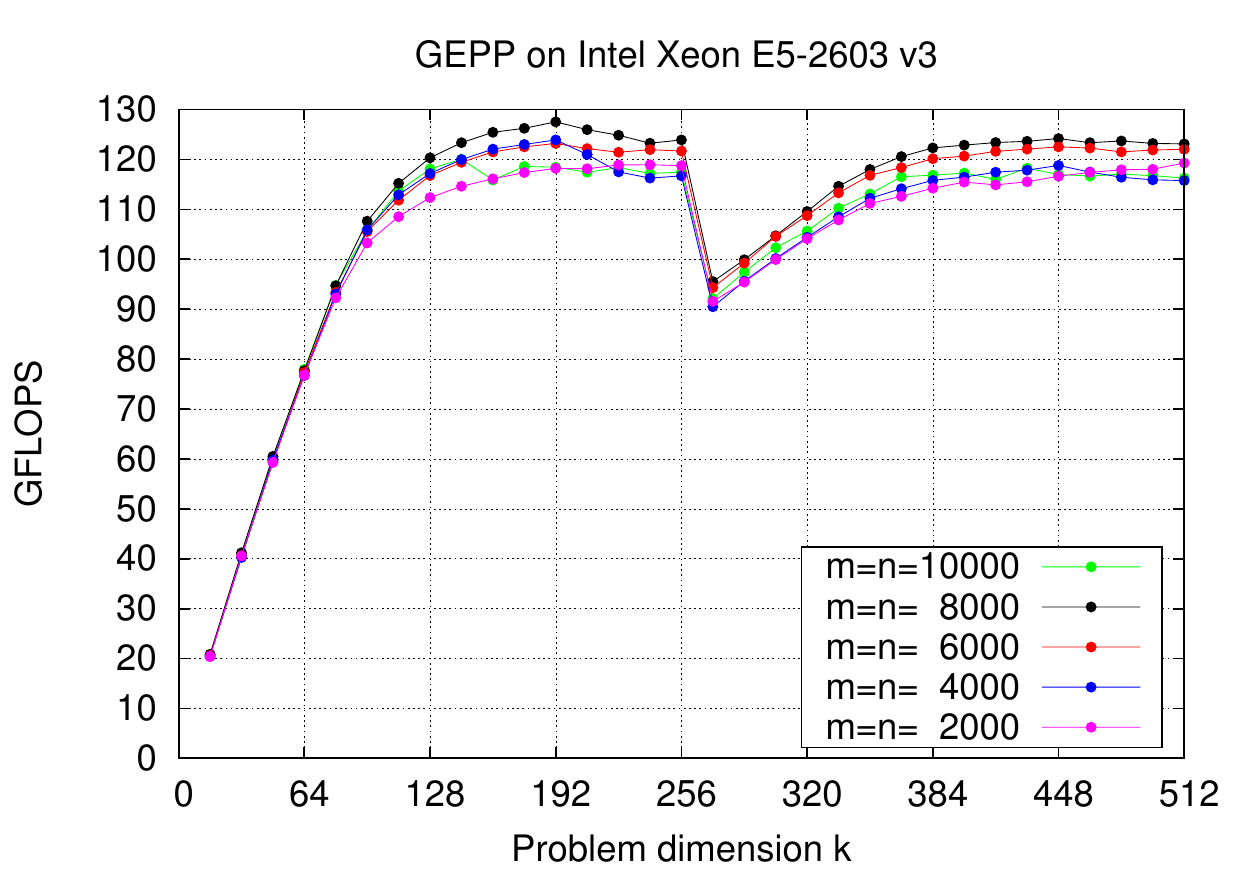}
\includegraphics[width=0.48\textwidth]{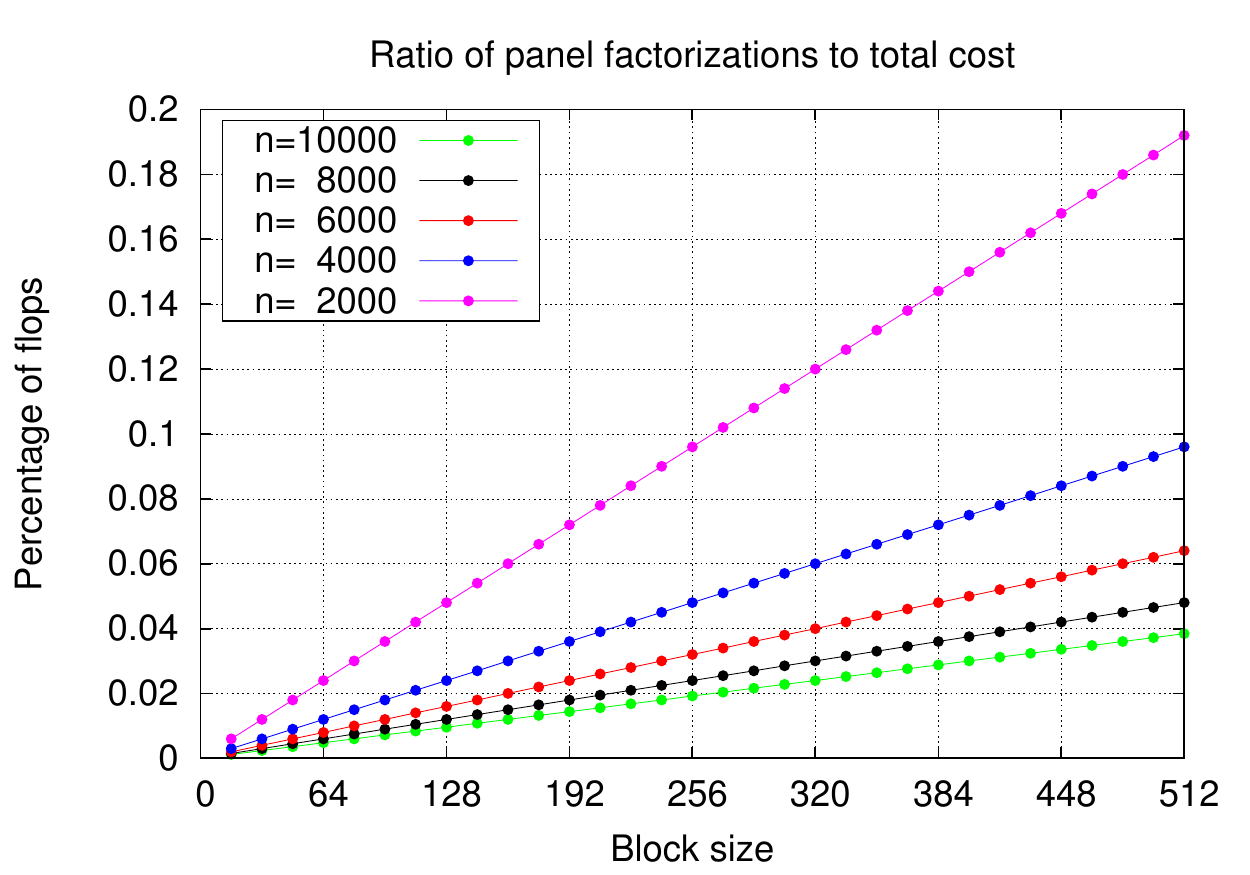}
\caption{GFLOPS attained with \gepp (left) and ratio of flops performed in the panel factorizations normalized to the total cost (right).}
  \label{fig:effect_block}
\end{figure*}

The performance of the blocked LU algorithms is strongly influenced by the outer block size $\bo$. As discussed
in subsection~\ref{subsec:basic}, this parameter should balance two criteria:
\begin{itemize}
\item Deliver high performance for the \gemm kernel. 
      Concretely, in the algorithms in Figures~\ref{fig:LU} and~\ref{fig:LU_la_blk}, 
       a value of $\bo$ that is too small turns 
      $A_{21}$ and $A_{12}$/$A_{12}^R$ into narrow column and row panels respectively, transforming the matrix multiplication involving
      these blocks ({\sf RL3} in Figure~\ref{fig:LU} and {\sf RU2} in Figure~\ref{fig:LU_la_blk}) into a memory-bound kernel that will generally deliver low performance.
      In the following, we will refer to a \gemm (\ref{eqn:gemm}) with dimensions $m \approx n\gg k$, as a panel-panel multiplication
      (\gepp)~\cite{Goto:2008:AHM:1356052.1356053}. Note that, for the \gepp arising in the LU factorizations, $k=\bo$.
\item Reduce the amount of operations performed in the panel factorization (about $n^2\bo/2$ flops, provided $n\gg \bo$), 
      in order to avoid the
      negative impact of this intrinsically sequential stage.
\end{itemize}
Figure~\ref{fig:effect_block} sheds further light on the roles played by these two factors.
The plot in the left-hand side reports the performance of \gepp, in terms of GFLOPS (billions of FLOPS),
showing that the implementation of this kernel in BLIS achieves an asymptotic performance peak for $k(=\bo)$ 
around~144.\footnote{The performance drop observed for $k$ slightly above 256 is due to the
optimal value of $k_c$ being equal to that number in this architecture.}
The right-hand side plot reports the ratio of flops performed in the panel factorizations with respect to
those of the LU factorization. %$3b_o/(4n)$ vs $2n^3/3$. 

\begin{figure*}[tb!]
\centering %
        \includegraphics[width=0.7\textwidth]{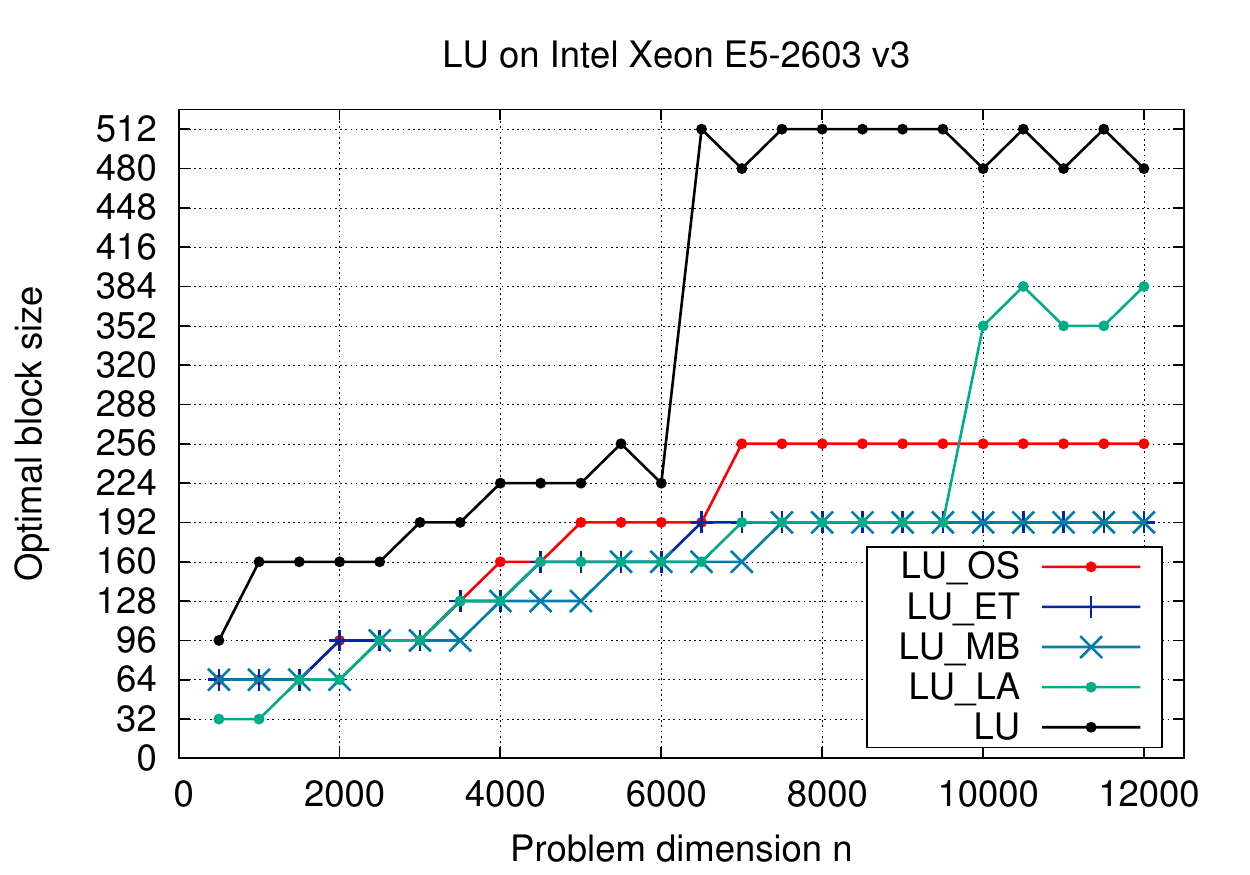}
\caption{Optimal block size of the blocked RL algorithms for the LU factorization.}
  \label{fig:optimal_block}
\end{figure*}

The combined effect of these criteria seems to point in the direction of choosing the smallest $\bo$ that 
attains the asymptotic GFLOPS rate for \gepp.
%Nevertheless, our results in 
However, Figure~\ref{fig:optimal_block} illustrates  
the experimental optimal block size $\bo$ 
for the distinct LU factorization algorithms,
exposing that this is not the case.
We next discuss the behavior for 
{\sf LU}, 
{\sf LU\_LA} and
{\sf LU\_MB}, which show different trends.
({\sf LU\_ET} and {\sf LU\_OS} are analyzed latter.)
In particular, {\sf LU} benefits from the use of larger values of $\bo$ than the other two
codes for all problem dimensions.
The reason is that a large block size operates on wide panels, which turns their factorization into a BLAS-3 operation with 
a mild degree of parallelism, and reduces the impact of this computation on the critical path of the factorization.
{\sf LU\_LA} exhibits a similar behavior for large problems, but favors smaller block sizes for
small to moderate problems. The reason is that, for {\sf LU\_LA},
it is important
to balance the panel factorization (\Tp) and remainder update (\Tr) so that their execution approximately requires the same time.

Compared with the previous two implementations,
{\sf LU\_MB} promotes the use of small block sizes, up to
$\bo$=192, for the largest problems. (Interestingly, this corresponds to the optimal value of $k$ for \gepp.)
One reason for this behavior is that, 
when the malleable version of BLIS is integrated into {\sf LU\_MB}, the practical costs of the two branches/tasks
do not need to be balanced.
Let us elaborate this case further, by considering the effect of reducing the block size, for example, from 
$\bo$ to $\bo'=\bo/2$. 
For simplicity, in the following discussion we will use approximations for the block dimensions and their costs;
furthermore, we will assume that $n\gg \bo$.
The first and most straight-forward consequence of halving the block size is that the number of iterations is doubled.
Inside each iteration with the original block size $\bo$, the loop body invokes, among others kernels,
a \gepp of 
dimensions $m\times (m-\bo)\times \bo$ 
(with $m$ the number of rows in the trailing submatrix $A_{22}^R$), for a cost of $2m^2\bo$ flops;
in parallel, the factorization involves a panel of dimension $m\times \bo$, for a cost of $m\bo^2-\bo^3/3\approx m\bo^2$ flops.
When the block size is halved to $\bo'$, the same work is basically computed in two consecutive iterations. However,
this reduces the amount of flops performed in terms of panel factorizations to
about $2m(\bo')^2 = m\bo^2/2$ while it has a minor impact on the number of flops 
that are cast as \gepp (two of these products, at a 
cost of $2m^2\bo'=2m^2\bo/2$ flops each).
The conclusion is that, by reducing the block size, we decrease the time that the single thread spends in the panel factorization
\Tp, favoring its rapid merge with the thread team that performs the remainder update \Tr. Thus, in case the execution time of the LU
is dominated by \Tr, adding one more thread to perform this task (in this scenario, in the critical path) as soon as possible 
will reduce the global execution time of the algorithm.

\subsection{Performance comparison of the variants with static look-ahead}

The previous analysis on the effect of the block size exposes that choosing the optimal block size is
a difficult task. Either we need a model that can accurately predict the performance of each building
block appearing in the LU factorization, or we perform an extensive experimental analysis to select
the best value. The problem is even more complex if we consider that, in practice, an optimal selection would have
to vary the block size as the factorization progresses. Concretely, for the factorization of a square matrix
of order $n$ via a blocked algorithm, the problem is decomposed into multiple subproblems that involve the factorization
of matrices of orders $n-\bo$, $n-2\cdot \bo$, $n-3\cdot \bo$, etc. From Figure~\ref{fig:optimal_block}, it is clear
that the optimal value of $\bo$ will be different for several of these subproblems. In the end, the
value that we show in Figure~\ref{fig:optimal_block} for each problem 
has to be considered as a compromise that attains fair performance for a wide range of 
the subproblems appearing in that case.

\begin{figure*}[tb!]
\centering %
\includegraphics[width=0.7\textwidth]{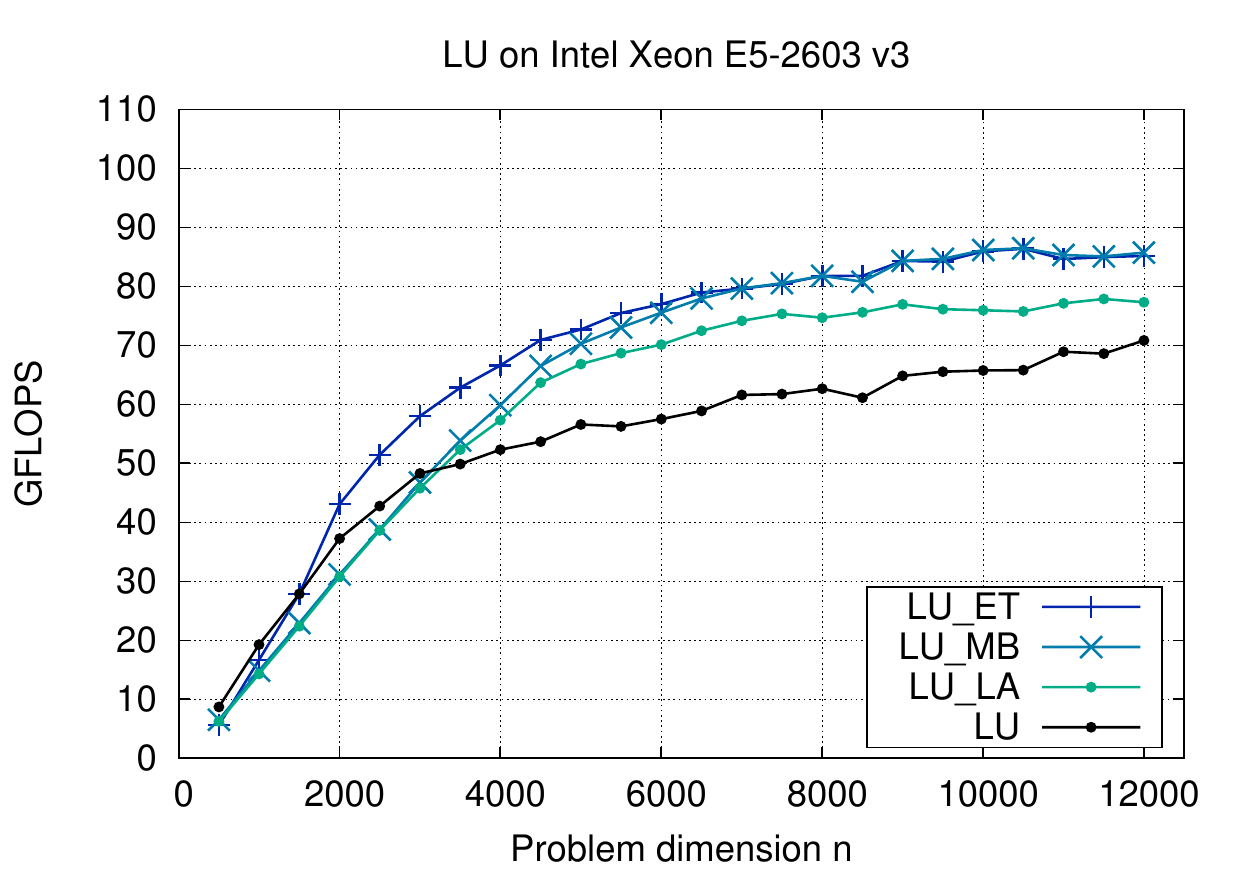}
\caption{Performance comparison of the blocked RL algorithms for the LU factorization (except {\sf LU\_OS})
         with a fixed block size $\bo=256$.}
  \label{fig:static}
\end{figure*}

Figure~\ref{fig:static} 
reports the GFLOPS rates attained by the distinct implementations to compute the plain
LU factorization and the variants equipped with static 
look-ahead (i.e., all except {\sf LU\_OS}), using $\bo=256$ as a compromise value for all of them.
Although this value is optimal for only a few cases, the purpose of this experiment is to show the improvements
attained by gradually introducing the techniques enhancing look-ahead.
The figure reveals some relevant trends:
\begin{itemize}
\item Except for the smallest problems, integrating the look-ahead techniques clearly
      improves the performance of the plain LU factorization implemented in {\sf LU}. 
\item The version with malleable BLAS ({\sf LU\_MB}) improves the performance of the basic version of look-ahead ({\sf LU\_LA})
      for the larger problems.  This is a consequence of the cost
      of the panel factorization relative to that of the global factorization. Concretely, 
      for fixed $\bo$,
      as the problem size grows, the global flop-cost varies cubically in $n$, as $2n^3/3$, 
      while the flop-cost of the panel factorizations grows quadratically, with $n^2\bo/2$.
      Thus, we can expect that, for large $n$, the remainder update \Tr becomes more expensive than the panel factorization
      \Tp. This represents the actual scenario that was targeted by the variant with malleable BLIS.
\item The version that combines the malleable BLAS with ET ({\sf LU\_ET}) 
      delivers the same performance of {\sf LU\_MB} for large problems, but outperforms
      all other variants with static look-ahead for the smaller problems. Again, this could be expected by considering
      the relative cost of the panel factorization for small $n$.
\end{itemize}

\subsection{Performance comparison with OmpSs}

We conclude the experimental analysis by providing a comparison of the best variant with static look-ahead,
{\sf LU\_ET}, with the implementation that extracts parallelism via the OmpSs runtime, {\sf LU\_OS}.
In this last experiment we depart from the previous case, performing an extensive evaluation
in order report the performance for the optimal block size for each problem dimension and algorithm. 
(See Figure~\ref{fig:optimal_block} for the actual optimal values employed in the experiment.)
For {\sf LU\_OS}, we select a value for $\bo$ that is then fixed for the complete factorization. 
As this variant overlaps the execution of tasks from different iterations in time, it is difficult
to vary the block size as the factorization progresses.
For {\sf LU\_ET}, the selected value of $\bo$ only applies to the first factorization. After that, 
the ET mechanism automatically adjusts this value during the iteration.

\begin{figure*}[tb!]
\centering %
        \includegraphics[width=0.7\textwidth]{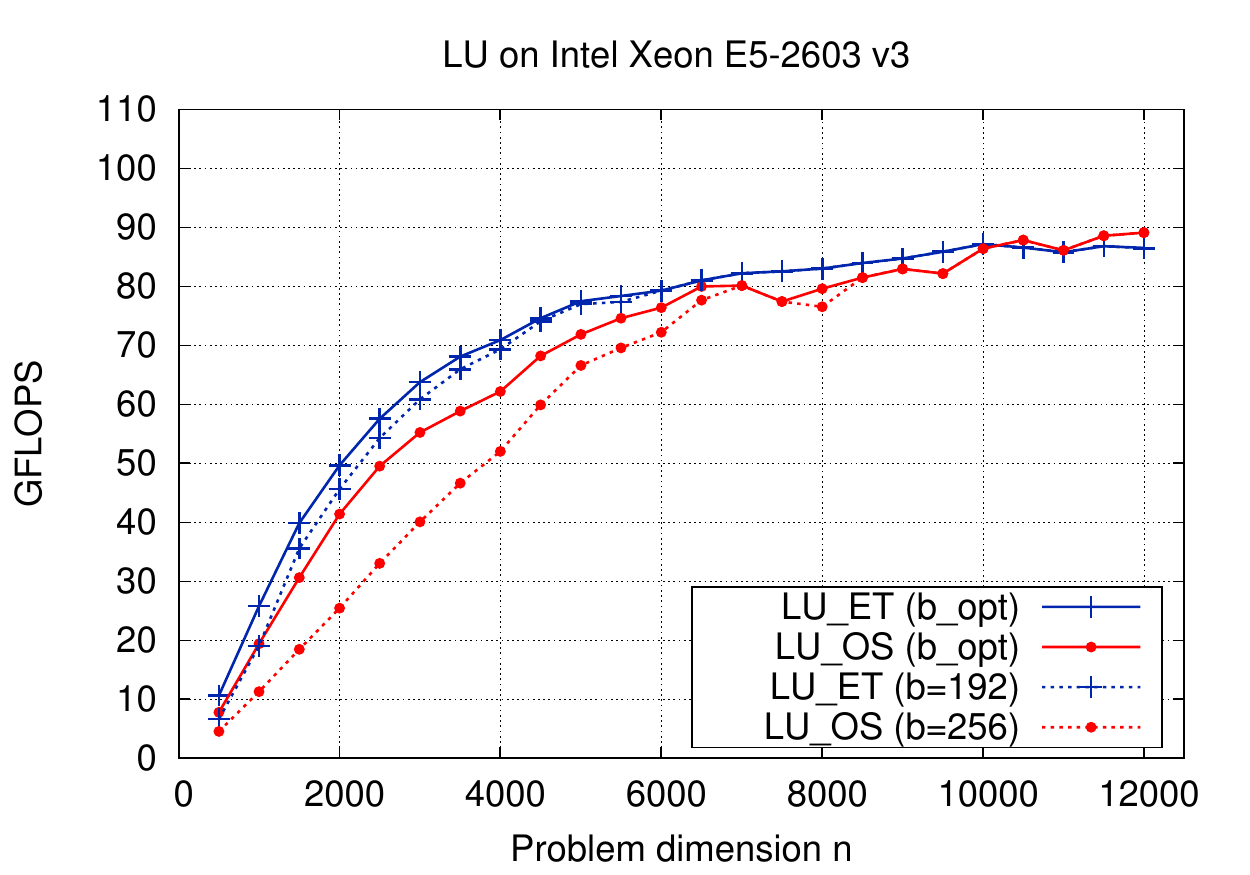}
\caption{Performance comparison between the OmpSs implementation and the blocked RL algorithm for the LU factorization 
         with look-ahead, malleable BLIS and ET.
         Two configurations are chosen for each algorithm: optimal block size for each problem size; and fixed block sizes $\bo=192$ for {\sf LU\_ET} and $\bo=256$ for {\sf LU\_OS}.}
  \label{fig:dynamic}
\end{figure*}

Figure~\ref{fig:dynamic} shows the results for this comparison in the lines
labelled as ``{\sf (b\_opt)}''.
{\sf LU\_ET} is very competitive, clearly outperforming the runtime-based solution for 
most problems and offering competitive performance for the largest three.

Manually tuning the block size to each problem dimension is in general impractical.
For this reason, the figure also shows the performance curves when the block size is fixed
to $\bo=192$ for {\sf LU\_ET} and 
$\bo=256$ for {\sf LU\_OS}. These values were selected because they 
offered high performance for a wide range of problem dimensions
(especially, the largest ones; see Figure~\ref{fig:optimal_block}). 
Interestingly, the performance lines corresponding to this configuration, labelled
with ``{\sf (b=192)}''/``{\sf (b=256)}'', show that choosing a suboptimal value for
$\bo$ has a minor impact on the performance of our solution {\sf LU\_ET}, because the
ET mechanism adjust this value on-the-fly (for the smaller problem sizes). Compared with this,
the negative effect of a suboptimal selection on {\sf LU\_OS} is clearly more visible.

A comparison with other parallel versions of the LU factorization with partial pivoting is possible,
but we do not expect the results change the message of our paper.
In particular, Intel MKL includes a highly-tuned routine for this factorization that relies in their
own implementation of the BLAS and some type of look-ahead.  
Therefore, whether the advantages of one implementation
over the other come simply from the use of a different version of the BLAS, or from the positive effects of our 
WS and ET mechanism, will be really difficult to infer. 
%because of the distinct versions of the BLAS underlying the two implementations (BLIS vs Intel MKL), and
%the impossibility of inspecting/changing Intel's.
The PLASMA library~\cite{plasmaweb} also provides a routine for the LU factorization with partial pivoting supported
by a runtime that implements dynamic look-ahead. The techniques integrated in PLASMA's routine are not different
from those in the OmpSs implementation evaluated in our paper. Therefore, when linked with BLIS, we do not expect
a different behaviour between PLASMA's routine and {\sf LU\_OS}.

\section{Concluding Remarks and Future Work}
\label{sec:remarks}

%We have implemented and evaluated several codes for the LU factorization with partial pivoting on 
%current multicore processors. For the version of this factorization equipped with static look-ahead, we have
%described how
%to tackle an unbalanced distribution of the workload between the panel factorization \Tp and the update \Tr during the factorization. 
%Concretely, for those iterations where \Tp is more expensive than \Tr, we propose to leverage our own malleable thread-level implementation of BLIS, which offers an entry point that allows threads to unite the execution of a BLAS kernel that is on-the-fly.
%Otherwise, in case \Tr more expensive than \Tp, 
%we enforce an early termination of the panel factorization, which promotes all threads into the execution of the next 
%iteration to avoid idle periods.
We have introduced WS and ET as two novel techniques to avoid workload imbalance during the execution of
matrix factorizations, enhanced with look-ahead, for the solution of linear systems.
The WS mechanism especially benefits from the adoption of a malleable thread-level instance of BLIS,
which allows the thread team in charge of the panel factorization, upon completion of this task, to be reallocated to the execution of the trailing update.
The ET mechanism tackles the opposite situation, with a panel factorization that is costlier than the trailing update. In such scenario,
the team that performed the update communicates to the second team that it should terminate the panel factorization, advancing
the factorization process into the next iteration.

Our results on an Intel Xeon E5-2603 v3 show the performance benefits of our version enhanced with 
malleable BLIS and ET compared with a plain LU factorization as well as a version with look-ahead.
The experiments also report competitive performance compared with an LU factorization that is parallelized by
means of a sophisticated runtime, such as OmpSs, that introduces look-ahead of dynamic (variable) depth.
Compared with the OmpSs solution, our approach offers higher performance for most 
problem dimensions, seamlessly tunes the algorithmic block size, and features a considerably smaller 
memory footprint as it does not require a sophisticated runtime support.

To conclude, our paper does not intend to propose an alternative to runtime-based solutions.
Instead, the message implicitly carried in our experiments aims to emphasize the benefits of 
malleable thread-level libraries, which we expect to be crucial 
in order to exploit the massive thread parallelism of future
architectures.
This work opens a plethora of interesting questions for future research. In particular,
how to generalize the ideas to a multi-task scenario,  what kind of interfaces may ease thread-level malleability,
and what kind of support is necessary in the runtime for this purpose.
%Static look-ahead is the way to go for MPI. Our ideas should be portable to the MPI world!

\subsection*{Acknowledgements}
We thank the other members of the FLAME team for their support.
This research  was partially sponsored by  projects TIN2014-53495-R and 
TIN2015-65316-P  of  the Spanish  {\em  Ministerio  de Econom\'{\i}a  y 
Competitividad}, the H2020 EU  FETHPC Project 671602 ``INTERTWinE'', by 
project 2014-SGR-1051 from the Generalitat  de Catalunya, and NSF grant 
ACI-1550493.                                                            

{\em Any opinions, findings and conclusions or recommendations
expressed in this material are those of the author(s) and do not
necessarily reflect the views of the National Science Foundation
(NSF).}

%\begin{thebibliography}{00}
%
%%% \bibitem{label}
%%% Text of bibliographic item
%
%\bibitem{}
%
%\end{thebibliography}

%\bibliographystyle{elsarticle-num}
%\bibliography{biblio}

\end{document}